\documentclass[twocolumn]{revtex4-1}

\usepackage[T1]{fontenc}
\usepackage{graphicx}

\usepackage{amsmath}
\usepackage{times}
\usepackage{mhchem}
\usepackage{multirow}
\usepackage{siunitx}

\newcommand{\gowo}{$G_0W_0$}
\newcommand{\gwo}{$GW_0$}
\newcommand{\hzos}{(Hf,Zr)O$_2$}
\newcommand{\hzo}{HZO}
\newcommand{\hzohalf}{Hf$_{0.5}$Zr$_{0.5}$O$_2$}

\begin{document}

\title{Structural and electronic properties of (Hf,Zr)O$_2$ phases from first principles}
\author{Sibghatullah Khan}
\author{Hannu-Pekka Komsa}
\email{hannu-pekka.komsa@oulu.fi}
\affiliation{%
Advanced Electronics Centre, Faculty of Information Technology and Electrical Engineering, University of Oulu, P.O. Box 4500, Oulu, FIN-90014, Finland
}%

\date{\today}

\begin{abstract}
(Hf,Zr)O$_2$ material system is promising for ferroelectric devices, owing to their compatibility with the existing electronics manufacturing processes. 
Technology computer-aided design (TCAD) simulations can be used for the device design and optimization, but the predictive power of TCAD device simulations depends on the quality of the material parameters. These are still surprisingly poorly understood, also for the metastable ferroelectric phase, largely owing to the difficulties in synthesizing phase-pure samples. 
Here, we carry out high-quality first-principles calculations to extract reliable lattice parameters, dielectric constants, and band gaps for six of the most common phases of \ce{HfO2}, {\hzohalf}, and \ce{ZrO2}.
We compare results from GGA, meta-GGA, and hybrid functionals to shed light on the sensitivity of the results on the level of theory and to connect the results to earlier literature. In the case of band structures, we also consider $GW$ corrections as well as the effects of temperature-induced band gap renormalization and excitonic effects. Finally, we use these results in TCAD simulations for (Hf,Zr)O$_2$ FeCAP to demonstrate the effect that non-ferroelectric interlayers could have on the hysteresis curves and band diagrams.
\end{abstract}

\maketitle


\section{Introduction}

A large portion of energy in modern chips is consumed in moving data between memory and the processor cores \cite{horowitz20141,sebastian2020memory}. 
Reduction of the energy consumption in future electronic devices requires new computer architecture paradigms, where the memory and information processing capabilities are intertwined more closely \cite{sebastian2020memory}.
One promising solution is introducing ferroelectricity into the components to provide persistent (non-volatile) memory state in a way that can be directly integrated to the electronic components, such as by changing a conventional (dielectric) gate oxide to a ferroelectric material in a field-effect transistor \cite{yu2025ferroelectric,mikolajick2021next,aziz2018computing}.
This, however, requires materials that are compatible with the chip manufacturing processes.
Currently, {\hzos} system, and the equimolar composition {\hzohalf} (HZO) in particular, is perhaps the leading candidate for this owing to its low growth temperature below 450°C, required for back-end-of-line compatibility, and vast previous experience with \ce{HfO2} as high-k gate oxide \cite{boscke2011ferroelectricity,muller2013ferroelectric,hwang2025recent}.
 
Despite the high technological interest in {\hzo}, some material properties are still surprisingly poorly understood.
{\hzo} can exist in several phases (polymorphs), but synthesis of phase-pure material is challenging, which hinders determination of the phase-specific properties \cite{Muller12_NL,zarubin2016fully}. 
Moreover, much of the recent experimental research has focused on the ultrathin films that exhibit ferroelectricity, which pose further challenges in the material characterization:
(i) Ferroelectricity is present in only some of the metastable phases \cite{park2015ferroelectricity,Zhu24_npjCM},
(ii) the low-temperature synthesis results in films with small grains of multiple phases \cite{park2017surface}, and
(iii) optimal ferroelectric properties are achieved at about 10 nm thickness \cite{kim2018effect}.
Consequently, significant effort has been devoted to understand and optimize growth processes that can maximize the content of ferroelectric phase and the ferroelectric properties, but usually involving non-equilibrium growth conditions and strong dependence on the surrounding materials \cite{mikolajick2021next,park2015ferroelectricity}.
For rational material and device optimization, especially when involving device simulations, access to accurate material properties is critically important.

First-principles calculations can be used to evaluate the missing material properties.
\ce{HfO2} and \ce{ZrO2} properties have been calculated in numerous papers \cite{Jaffe05_PRB,Fadda10_PRB,Jiang10_PRB,Gruning10_PRB,ReyesLillo14_PRB,Materlik15_JAP,Sklenard18_APL,Kersch22_FN,AzevedoAntunes22_PSS,Zhu24_npjCM}.
However, these works are still lacking in several aspects.
First, most of them are carried out only at the (semi-)local level \cite{Jaffe05_PRB,Fadda10_PRB,ReyesLillo14_PRB,Materlik15_JAP,Kersch22_FN,AzevedoAntunes22_PSS,Zhu24_npjCM}, i.e., employing LDA or PBE functionals, which may not be sufficiently accurate.
These functionals are at the lowest levels of the ``Jacob's ladder'' of functionals \cite{Perdew01}, to be followed by meta-GGA (dependence on kinetic energy density), hybrid functionals (dependence on occupied orbitals), and finally RPA/GW (dependence on unoccupied orbitals).
Second, some studies have not included all the relevant low-energy phases.
Importantly, the non-polar orthorhombic phase has not been included in many older studies \cite{Jaffe05_PRB,Fadda10_PRB,ReyesLillo14_PRB,Materlik15_JAP}, although it has the lowest energy of the metastable phases \cite{Kersch22_FN,AzevedoAntunes22_PSS,Zhu24_npjCM}.
Finally, explicit models for {\hzos} are fairly rarely considered in such studies.

In this paper, we report first-principles calculations of structural and electronic properties of six of the most relevant phases of \ce{HfO2}, \ce{ZrO2}, and {\hzohalf} carried out at increasing levels of theory.
We first compare semilocal functionals to meta-GGA and hybrid functionals, and show that there is a consistent improvement in lattice parameters, formation energies, and dielectric constants when moving to the higher rungs of the ladder.
In the case of band structures, we also carry out GW calculations and evaluate temperature-induced band gap renormalization and excitonic effects.
Throughout the paper, we also show the results for \ce{HfO2} and \ce{ZrO2}, in addition to {\hzo}, since
(i) they are much better characterized and can thus be used for benchmarking computational methods,
(ii) the Hf/Zr ratio does not need to be 0.5 and using the values for the end points, other values can be interpolated,
(iii) to provide comprehensive set of properties for phases other than monoclinic, which may be missing in the literature, and (iv) \ce{HfO2} and \ce{ZrO2} are also technologically important materials, even if not for ferroelectric devices.

\section{Computational details}

\begin{figure*}[ht]
\begin{center}
  \includegraphics[width=12cm]{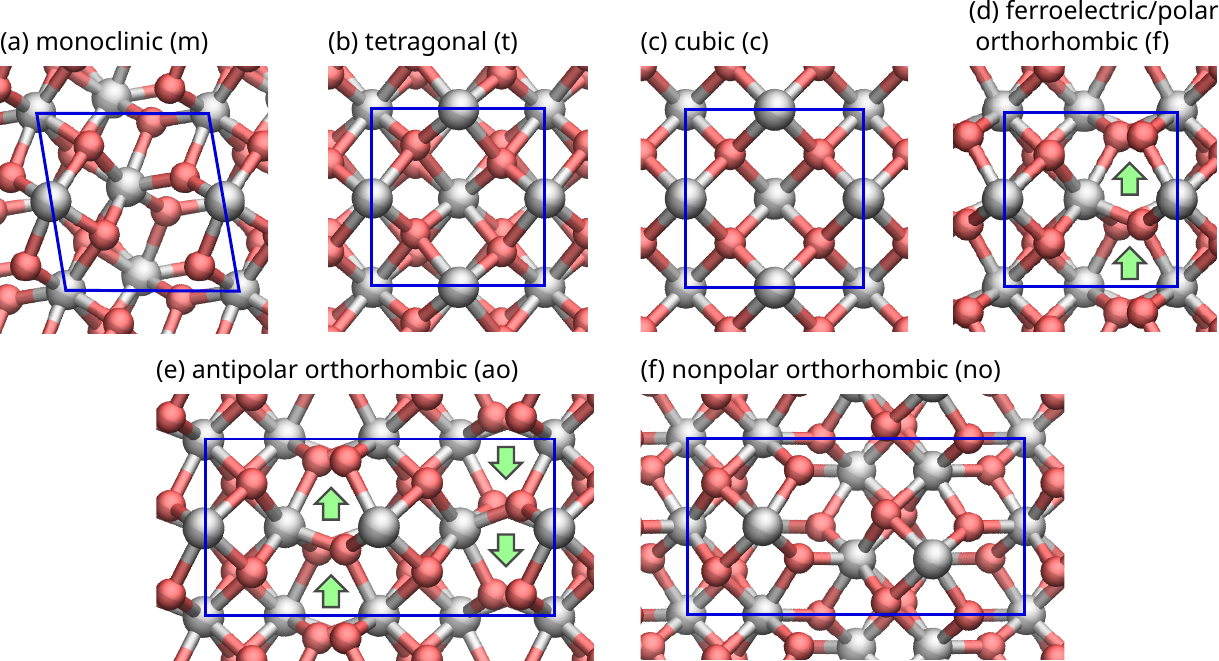}
\end{center}
\caption{\label{fig:geos}
  Atomic structures of all considered phases, taken from HSE06 calculations for \ce{ZrO2}. \ce{HfO2} and {\hzo} structures are very similar.
  Silver spheres are Zr(Hf) atoms and red spheres O atoms.
  Blue rectangle/rhomboid denotes the unit cell. The O displacements relevant for the polarization are indicated with light green arrows.
}
\end{figure*}

We here consider six phases for all three materials:
monoclinic (P2$_1$/c), tetragonal (P4$_2$/nmc), cubic (Fm$\bar{3}$m), ferroelectric/polar orthorhombic (Pca2$_1$), antipolar orthorhombic (Pbca), and nonpolar orthorhombic (Pbca).
In the following, these are denoted as m-phase, t-phase, c-phase, f-phase, ao-phase, and no-phase.
The atomic structures for these phases are illustrated in Fig.~\ref{fig:geos}.
These six phases are those commonly observed in experiments and also exhibit the lowest calculated formation energy \cite{Zhu24_npjCM}.
Consequently, we are ignoring few of the higher energy phases proposed in the literature \cite{Zhu24_npjCM}.

The atomic structures used as a starting point were taken from experimentally reported phases for \ce{ZrO2}. To guarantee consistent unit cell choice, the same structures were also used for \ce{HfO2} and {\hzo} after substituting the metal atoms. We tested several different orderings for Hf and Zr in {\hzo} and selected the lowest energy structure, although the energy variation for different ordering of Hf and Zr were very small, less than 1 meV/formula unit.

All density-functional theory calculations were carried out within the projector-augmented wave framework as implemented in VASP \cite{kres3,kres4}.
We benchmark PBE\cite{Perdew96}, r2SCAN\cite{Furness20_JPCL}, and HSE06\cite{Heyd03,Heyd06} exchange-correlation functionals.
In all calculations, we use plane-wave cutoff of 500 eV and PAW setups with 12 valence electrons for Hf and Zr and 6 valence electrons for O (\verb|Hf_sv|, \verb|Zr_sv| and \verb|O| in VASP 5.4 PAW set).
Most calculations adopted the 12-atom conventional unit cell, or the 24-atom unit cell in the case of ao- and no-phases. For these, we used 6x6x6 and 6x6x3 k-point meshes, respectively. Selected calculations were carried out using the primitive cells, 3-atom cell for c-phase and 6-atom cell for t-phase of \ce{HfO2} and \ce{ZrO2}, with 8x8x8 k-point mesh.
Atomic structures were optimized until forces were below 1 meV/{\AA}.
In hybrid functional calculations, we reduced the k-point mesh for describing the Fock exchange potential by a factor of two and used a reduced FFT grid.
To enable Fock mesh reduction in the case of ao- and no-phases, their k-point meshes were increased to 6x6x4.
The x-ray diffraction (XRD) spectra were simulated with pymatgen \cite{pymatgen}, assuming Cu K$\alpha$ line.

The dielectric constants were calculated from a self-consistent response to finite electric field within modern theory of polarization \cite{Nunes01_PRB,KingSmith93_PRB}.
The electric field strength was set to 0.005 V/{\AA}. To guarantee highly converged wave functions, the SCF loop convergence criterion was set to $10^{-8}$ eV. 
The force constants (and vibrational frequencies) were calculated using finite difference accounting for the symmetry with atomic displacement of 0.015 {\AA}.


The parameters for $GW$ were carefully benchmarked, as shown in Fig.~S1.
We used same plane-wave energy cutoff and k-point mesh, and same k-point mesh reduction and FFT grid reduction for Fock potential as in HSE06 calculations.
Energy cutoff for the response function was set to 270 eV and the number of bands in the calculations to 384 per formula unit.
We also carried out partially self-consistent \gwo{} calculations by iterating eigenvalues in $G$, while keeping orbitals and $W$ fixed.

Bethe-Salpeter equation (BSE) calculations for c-phase were evaluated on top of PBE+$GW$ band structure, using the same 8x8x8 k-point mesh but without k-point reduction. Furthermore, we adopted the Tamm-Dancoff approximation and used 8 occupied and 8 virtual states.

In all cases, the band structures are obtained by interpolation using BoltzTrap2 package \cite{Boltztrap2}.
Temperature effects on the band structure were evaluated at the PBE level using the one-shot method \cite{Zacharias16_PRB}. The supercell size was 3x3x3 and k-point mesh 2x2x2.

TCAD simulations were carried out using Sentaurus to model the ferroelectric capacitor behavior \cite{chandra2007landau, ng2012depletion}. The device geometry consisted of a ferroelectric film with a nominal thickness of 10 nm, and compared to cases with 6 nm ferroelectric layer and a 4 nm non-ferroelectric layer.
These are sandwiched between two TiN electrodes, with a device width of 100 nm. A two-dimensional mesh was applied, with refinement at material interfaces, to accurately resolve electric field variations and polarization gradients. The ferroelectric layer was described using the Ginzburg--Landau--Khalatnikov (GLK) model with parameters $P_r = 35$ {\textmu}C/cm$^2$, $E_c = 3\, $MV/cm, $\rho = 2.25 \cdot 10^4 \, \Omega\cdot\text{cm}$ and $g = 1 \times 10^{-6}\, $cm$^3$/F taken from our previous work \cite{khan2026physically}. 
Permittivity and band gaps will be extracted from the first-principles results.
Hysteresis loops were obtained by applying a triangular voltage waveform to the top electrode while the bottom electrode was grounded, sweeping the applied voltage as 0 $\rightarrow$ +V$_{pp}$/2 $\rightarrow$ $-$V$_{pp}$/2 $\rightarrow$ 0 V. The voltage rise and fall times were set to $\tau = 125$ {\textmu}s with inter-pulse delays of $2\tau = 250$ {\textmu}s. The polarization was extracted directly as the volume-average value over the ferroelectric regions in each simulation step, capturing both switching and non-switching contributions throughout the complete cycle.

\section{Results}

\subsection{Lattice parameters}

We start by comparing the calculated structural parameters to the experimental ones.
These are fairly straightforward to extract experimentally and known for most
of the phases considered in this work, thus providing a good starting point for evaluating accuracy of the functionals.

\begin{figure*}[ht]
\begin{center}
  \includegraphics[width=17cm]{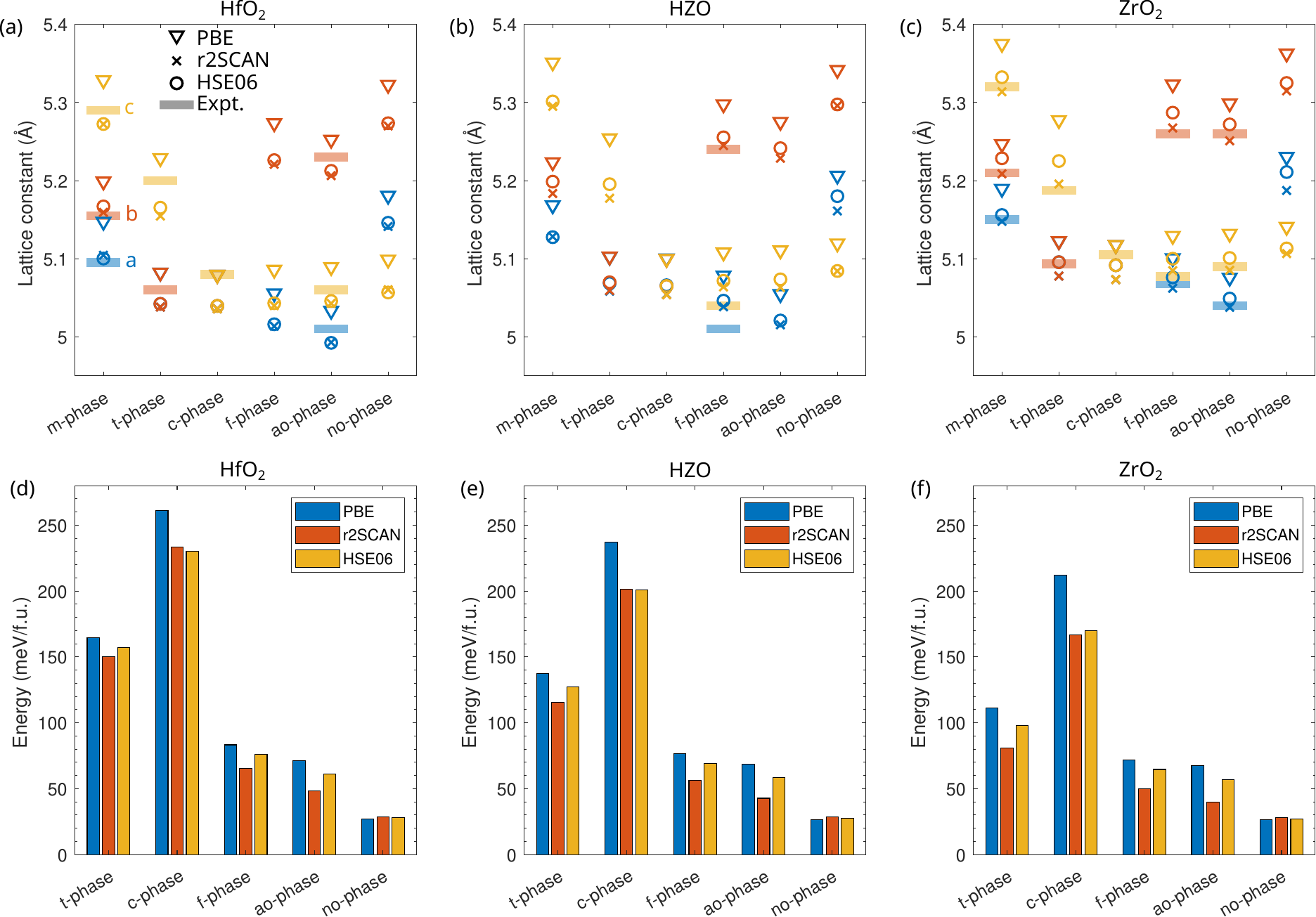}
\end{center}
\caption{\label{fig:latE}
(a-c) Lattice constants calculated using PBE, r2SCAN, and HSE06 functionals and comparison to the experimental values where available.
(d-e) Total energy differences (in meV/f.u.) with respect to the monoclinic phase.
}
\end{figure*}

The lattice constants from the three functionals in comparison to the experimental values (averaged over available ones, where available) are shown in Fig.~\ref{fig:latE}(a-c). 
All lattice constants are given in terms of the corresponding 12-atom conventional unit cell, i.e., the long lattice constant of ao- and no-phases has been divided by two.
Full data, including all experimental values and comparison to other calculations, can be found in Tables S1--S3.

PBE is known to usually underbind materials and consequently yield too large lattice parameters, while modern functionals tend to give improved lattice parameters\cite{Heyd05_JCP,Furness20_JPCL}.
These trends are also observed here. PBE yields clearly larger lattice constants than r2SCAN and HSE06, and larger than the experimental values.
r2SCAN values are consistently slightly lower than HSE06 ones, but both yield similarly good agreement with the experiments.
Only in the case of c-phase, PBE appears to match better with experiments,
although we note that experimental determination of c-phase lattice parameters at room temperature can be challenging due to its metastability.

Our results are in good agreement with previous PBE calculations
\cite{Zeng14_AC, Fadda10_PRB, Zhu24_npjCM}
and qualitatively with calculations using other GGA functionals \cite{Jaffe05_PRB,Huan14_PRB,Kersch22_FN,AzevedoAntunes22_PSS} and even LDA calculations \cite{Fadda10_PRB,ReyesLillo14_PRB,Materlik15_JAP}.
In early study by Materlik et al., LDA was found to give good lattice parameters for m-phase, underestimate for c- and t-phases, and clearly overestimate for o-phase \cite{Materlik15_JAP} (cf.\ Tables S1--S3).
The latter is particularly surprising, given that LDA usually leads to underestimation of lattice parameters.
Nevertheless, modern functionals give more balanced performance, yielding reasonable agreement with experiments in all cases.



The total energy differences with respect to the most stable monoclinic phase
are shown in Fig.~\ref{fig:latE}(d-f) and also included in Tables S1--S3.
Since these are energy differences, one cannot directly relate smaller or larger values to under- or overbinding, and consequently the trends between functionals are much less obvious: generally PBE yields the highest formation energy difference, r2SCAN the lowest, and HSE06 falling in between them,
but the differences are overall fairly small.
The LDA results from Ref.~\cite{Materlik15_JAP} (cf.\ Tables S1--S3) are consistently lower than those obtained by other functionals.
To the best of our knowledge, no experimental formation energy differences are available for these compounds.
Based on the earlier benchmarking studies, HSE06 and r2SCAN were found to give more accurate formation energies than PBE \cite{Heyd03,Paier06_JCP,Kothakonda23_ACSMau,Liu24_PRM},
and given that the r2SCAN and HSE06 give comparable results for all materials and phases (and lattice constants), we believe these results are more reliable than the PBE ones.


\begin{figure*}[ht]
\begin{center}
  \includegraphics[width=\textwidth]{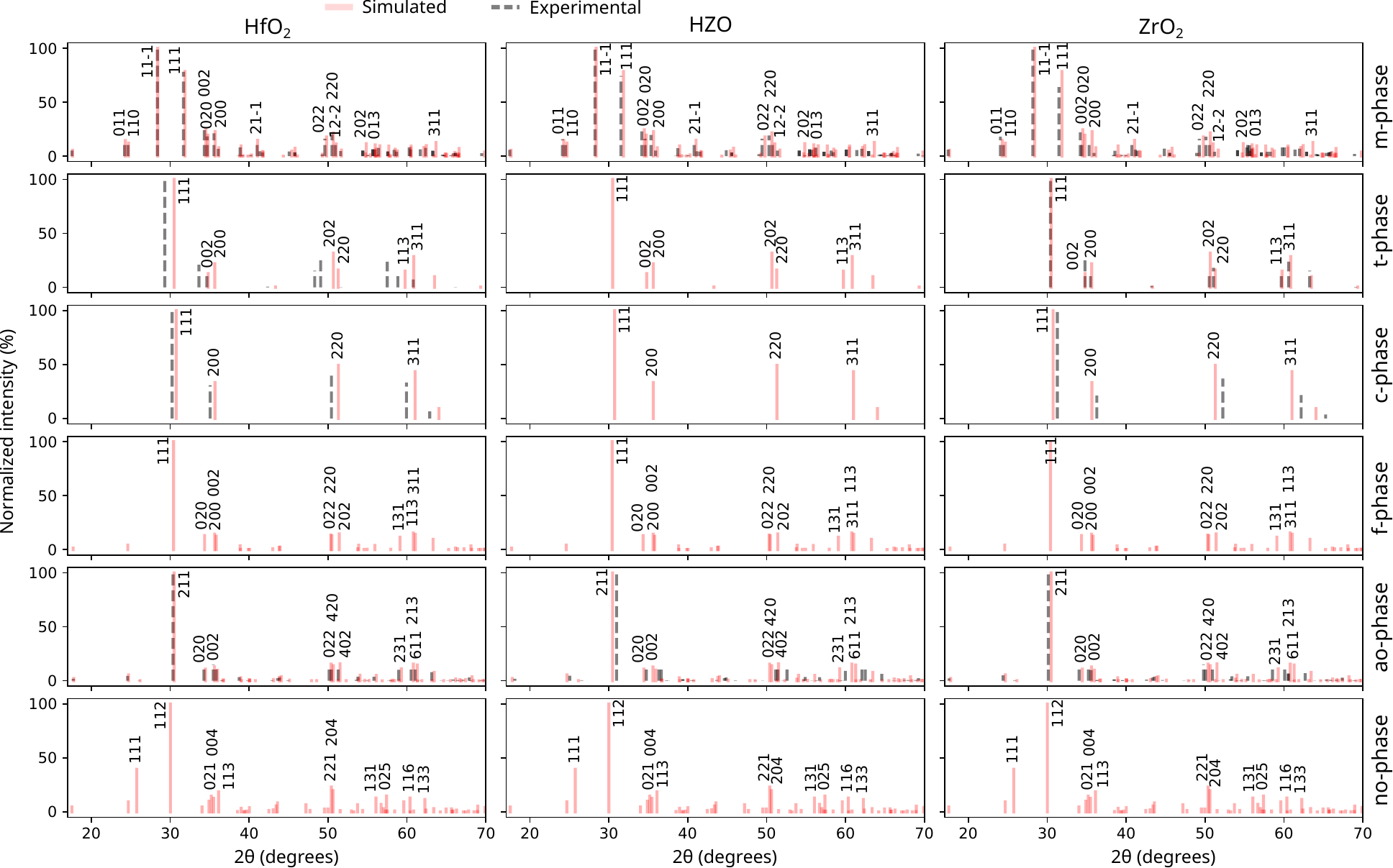}
\end{center}
\caption{\label{fig:xrd}
  XRD spectra of simulated (pink solid lines) and experimental (black dashed lines, where available) of all considered phases. The peaks are normalized (100\%) to the highest intensity peak. Only peaks with intensity over 10\% are labeled.
}
\end{figure*}

Finally, we simulated the powder XRD spectra of all phases and compared to the experiments.
In Fig.~\ref{fig:xrd}, we show the comparison using HSE-calculated crystal structures, which we assume to be the most accurate ones.
The indicated (hkl) reflections correspond to the unit cell definitions shown in Fig.~\ref{fig:geos}.
Experimental data are taken from PDF cards No. 01-089-9582, 04-002-5353, 04-002-0037, and 01-083-0808 for \ce{HfO2} \cite{Pathak20_ACC,Isupova68,Popov90,Ohtaka95_JACS}; No. 04-002-5428 and 04-027-2567 for {\hzo} \cite{Gavrish69,Johnson19_JAP}; No. 00-036-0420, 01-070-7303, 04-013-3441, and 04-007-0952 for \ce{ZrO2} \cite{Larson1985ICDD,Bouvier01_IJIM,Bouvier00_PRB,Ohtaka90}.
The experimental results are generally taken from the spectra measured at ambient conditions. The exceptions will be noted below.

Over all materials, the agreement is very good for m-phase and ao-phase, which are also the phases where the experiments are most reliable and converged. The simulations give slightly higher $2\theta$ (lower lattice constants), which can be explained by the finite-temperature effects increasing the experimental values, whereas the calculations predict 0 K results without zero-point energy.
It could also result simply from the inaccuracies of the calculations, but the results nevertheless demonstrate the good accuracy of the HSE-determined crystal structures.
ao-phase HZO would appear to be an exception with lower $2\theta$ from simulations, but the corresponding measurements were done at a pressure of 19.2 GPa, which explains the lower experimental lattice constants.

For the t- and c-phases, we selected the closest matches among several candidates with widely different spectra, but the agreement is still clearly worse.
Given the good agreement with HSE for m- and o-phases, it could mean that many of the reported entries of t- and c-phases are not phase-pure. 
We note that c-phase ZrO$_2$ was measured at a pressure of 28.9 GPa, but it gave significantly better agreement with simulations than other ambient condition results.

The simulations indicate that f-phase, and ao-phases have very similar main features and can be difficult to separate from experimental XRD. t-phase is also similar to these, unless the additional weak intensity features can be distinguished. m-phase, c-phase, and no-phase on the other hand have clear distinguishing features. 




\subsection{Dielectric constants}

\begin{figure*}[ht]
\begin{center}
  \includegraphics[width=17cm]{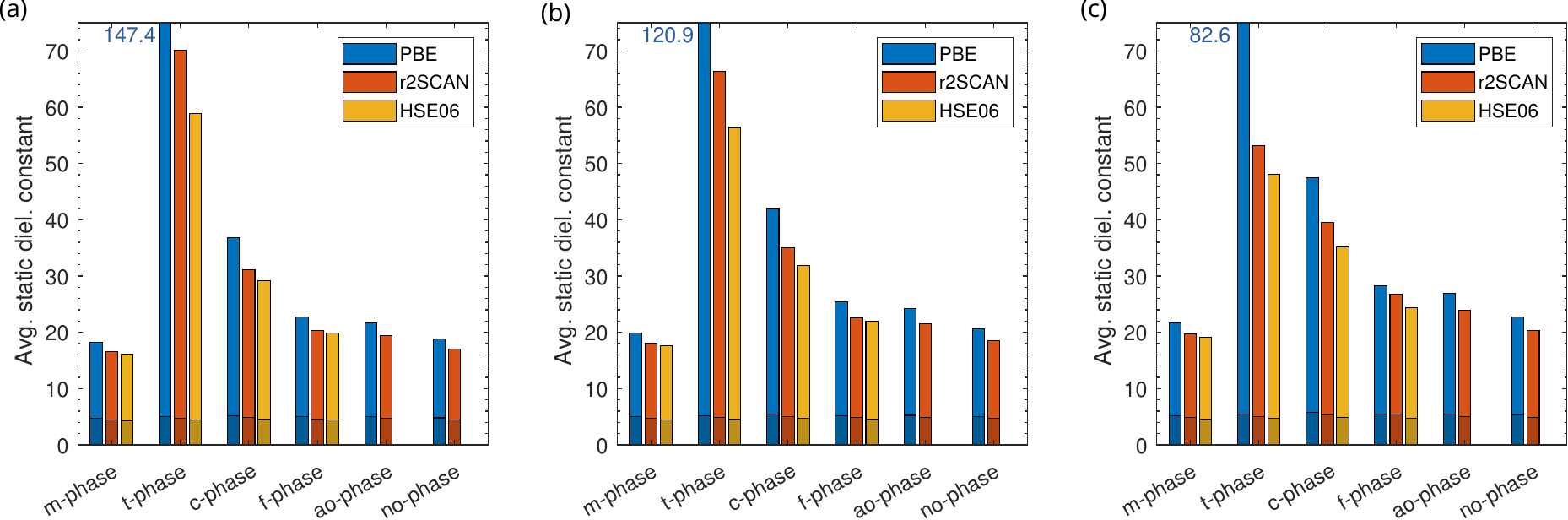}
\end{center}
\caption{\label{fig:alldiels}
Dielectric constants of all phases for (a) \ce{HfO2}, (b) {\hzo}, (c) \ce{ZrO2}.
The lighter and darker regions denote the ionic and electronic contributions to dielectric constant,
and the total bar height is the total static dielectric constant.
}
\end{figure*}


The calculated rotationally averaged static dielectric constants are shown in Fig.~\ref{fig:alldiels},
while the full numerical results are collected in Tables S4--S6.
A large variation between different phases is found for all materials,
with m-phase having the lowest dielectric constant and t-phase clearly the largest.
Moreover, t-phase dielectric constant is highly sensitive to the adopted functional.
For example, in the case of HZO, dielectric constants calculated using PBE, r2SCAN, and HSE06
functionals are 120.9, 66.4, and 56.4, respectively.
However, this variation arises purely from the ionic contribution, whereas the electronic
contribution is about 5 for all materials and all phases.
Several previous benchmarks have indicated that HSE06 should provide more accurate dielectric constants than PBE \cite{Paier08_PRB}, but we found only a small dataset benchmarking for r2SCAN dielectric constants \cite{Ning22_CM}.
Here, the very large value with PBE could arise from the aforementioned underbinding, which in turn means larger ionic response.
In similar vein, the LDA results of Materlik et al. do not suffer from same problem and are more similar to r2SCAN and HSE06 results \cite{Materlik15_JAP}.
For all phases except for tetragonal, the dielectric constants increase with increasing Zr content.
This can be ascribed to \ce{ZrO2} phases having slightly larger Born effective charges \cite{Materlik15_JAP,Bagheri23_SD} and/or their larger lattice parameters.

These results are in qualitative agreement with available experiments, but quantitative comparison is difficult due to the scarcity of the data and the experimental uncertainties.
To the best of our knowledge, dielectric constant determined from single-crystal sample is only available for m-phase \ce{ZrO2}, $22\pm 3$ found in Ref.~\cite{Harrop67_BJAP}.
The early review on high-k dielectric thin films by Wilk et al.\ listed a range of 22--25 for \ce{HfO2} and 14--25 for \ce{ZrO2} \cite{Wilk01_JAP}.
More recently, Tomida et al.\ reported 15 \cite{Tomida06_APL} and Martinez-Puente et al.\ 19 \cite{MartinezPuente22_MSEB} for m-phase \ce{HfO2}.
The calculated values with any of the adopted functionals fall within these ranges and
thus we cannot draw firm conclusions on which functional is the most accurate.

Studies of HZO films with different growth parameters have given insight into the evolution of dielectric properties with both Hf/Zr ratio and phase content.
Muller et al.\ reported an increase from $\varepsilon=23$ for mainly ($\sim$75\%) m-phase \ce{HfO2} to $\varepsilon=35$ for mainly t-phase \ce{ZrO2} \cite{Muller12_NL}.
Kim et al.\ reported an increase from $\varepsilon=14.6$ for mainly m-phase \ce{HfO2} to $\varepsilon=41$ for mainly t-phase Zr$_{0.6}$Hf$_{0.4}$O$_2$ \cite{Kim22_PSS}.
Ahn et al.\ reported an increase from $\varepsilon=23$ for mainly m-phase \ce{HfO2} to $\varepsilon=40$ for mainly t-phase \ce{ZrO2} \cite{Ahn15_ACSAMI}.
All of these results are in qualitative agreement about increasing dielectric constant with t-phase content.
Finally, Kashir et al. synthesized films with different ratio of tetragonal and orthorhombic phases and showed that the dielectric constant increases from around 30 at t/o=0.05 (mostly o-phase) to around 50 at t/o=0.4 \cite{Kashir21_PSSA}. A linear extrapolation of their results would yield 80 for pure t-phase.

Overall, the calculated m-phase and o-phase values still tend to be slightly smaller than in experiments,
whereas the calculated t-phase values tend to be larger than in experiments.
Part of the discrepancy could be explained by the anisotropy of the dielectric tensor.
For example, in r2SCAN calculations for HZO (Tables S4--S6), the dielectric constant varies from 15--20 for m-phase, 20--25 for f- and o-phase, and 18--97 for t-phase, depending on the direction.
Otherwise, the differences can arise from phase content, defects, and interface effects
such as unwanted interlayer phases.

The similarity of the calculated values for m- and o-phases with all functionals suggests that these are close to the true values for defect-free single-crystal materials.
As for the t-phase, given that there are no experimental reports reaching as high values as with PBE, it seems that PBE values are indeed overestimated,
and we would advise against using PBE for dielectric response calculations in these materials.



As a final remark, we note that since the dielectric response calculations yield also Born effective charges and the vibrational normal modes, the infrared spectra can be readily simulated. 
The results are collected in Fig. S2--S3 in SI, together with comparison to experiments. In line with other results herein, higher levels of theory are again found to yield more accurate vibrational properties. 
As Kersch et al. have already reported a detailed study of the vibrational properties of \ce{HfO2} and \ce{ZrO2}, we here refrain from further discussion and direct the reader therein.

\subsection{Band structure}

We next turn our attention to the band structure of these materials, where there is much less consensus, both experimentally and computationally.
Computationally, while (semi-)local can give fairly reliable atomic structure and formation energies, the band gaps are often seriously underestimated. Even hybrid functionals have limited predictive power. $GW$ calculations will improve the band gap values, but the errors can still be sizeable if only {\gowo} corrections are considered, that depend on the starting point.

Experimental challenges arise again from sample quality (phase purity and crystalline quality), indirect nature of the band gap, and simply variations arising from different measurement techniques (optical, electrical, etc.). 
Generally, reasonably reliable experimental data is only available for m-phase, varying 5.5--5.9 eV for \ce{HfO2} and 5.0--5.7 eV for \ce{ZrO2}, whereas very little data is available for other phases or for {\hzo}.
Therefore, in the following, we use the experimental data for m-phase for benchmarking the functionals and then use the most accurate functional to predict band gaps for other phases.







\begin{table*}[ht!]
\caption{\label{tab:gaps}
  Indirect and direct band gaps for m-phase calculated with different levels of theory
  and comparison to experimental values determined using different methods. 
  Electronic gap is obtained by summing up indirect HSE+\gowo{} band gap and temperature renormalization (T renorm. at 300 K).
  Optical gap is obtained by summing up direct HSE+\gowo{} band gap, temperature renormalization, and exciton binding energy ($E_b$ exciton, evaluated using c-phase). Electronic and optical gaps for {\hzo} are obtained by interpolation from \ce{HfO2} and \ce{ZrO2}.
}
\begin{tabular*}{\linewidth}{@{\extracolsep{\fill}} llcccccc}
& \multicolumn{2}{c}{\ce{HfO2}} & \multicolumn{2}{c}{\hzo} & \multicolumn{2}{c}{\ce{ZrO2}} \\
          & indir & dir & indir & dir & indir & dir \\
\hline
PBE      & 4.00 & 4.08 & 3.74 & 3.82 & 3.61 & 3.68 \\
r2SCAN   & 4.55 & 4.63 & 4.27 & 4.35 & 4.13 & 4.19 \\
HSE06    & 5.67 & 5.76 & 5.42 & 5.51 & 5.28 & 5.35 \\    
PBE+\gowo & 5.52 & 5.61 & 5.38 & 5.46 & 5.22 & 5.30 \\
PBE+\gwo  & 6.36 & 6.46 &      &      & 5.99 & 6.06 \\
HSE+\gowo & 6.27 & 6.37 &      &      & 5.93 & 6.01 \\
T renorm. (300 K) & -0.24 & -0.25 &       &       & -0.30 & -0.31 \\
Electronic gap & 6.03 &      & $\sim$5.83 &             & 5.63  & \\
$E_b$ exciton    &       & -0.34 &       &       &       & -0.31 \\
Optical gap    &      & 5.78 &            & $\sim$5.585 &       & 5.39\\
\hline
Expt. PES+IPS &
 \multicolumn{2}{c}{5.86\cite{Sayan04_JAP}, 5.7\cite{Bersch08_PRB}} & & & \multicolumn{2}{c}{5.68\cite{Sayan04_PSSB}, 5.5\cite{Bersch08_PRB}} \\
Expt. SE  &
 \multicolumn{2}{c}{5.56\cite{Nguyen04_JVSTA}, 5.5--6.0\cite{Hill08_JAP}}  & & & \multicolumn{2}{c}{5.25\cite{Zhu06_MSSP}} \\
Expt. EELS &
 \multicolumn{2}{c}{5.8\cite{Ikarashi03_JAP}, 5.3$\pm$0.5\cite{Cheynet07_JAP}}  & & & \multicolumn{2}{c}{5.50\cite{Miyazaki01_JVST}, 5.0\cite{Ikarashi03_JAP}} \\
Expt. other &  
 \multicolumn{2}{c}{5.70\cite{Puthenkovilakam04_JAP}, 5.65\cite{Balog77_TSF}}  & & & \multicolumn{2}{c}{5.65\cite{Puthenkovilakam04_JAP}} \\
\end{tabular*}
\end{table*}

Band gaps for m-phase calculated at different levels of theory are collected in Table \ref{tab:gaps}.
In addition to the three functionals considered previously, {\gowo} calculations evaluated on top of PBE or HSE06 results and {\gwo} on top of PBE are considered.
Our calculations for PBE and r2SCAN are generally in close agreement with previous calculations \cite{Ondracka16_JPD,Sklenard18_APL}.
The $GW$ results are found to vary more:
Jiang et al. reported PBE+{\gowo} and PBE+{\gwo} gaps of 5.45 and 5.78 for \ce{HfO2} and 4.99 and 5.34 eV for \ce{ZrO2} \cite{Jiang10_PRB}, respectively, in fair agreement with our results for {\gowo} (0.1--0.2 eV) but rather poor agreement for {\gwo} (0.6 eV).
Gr\"{u}ning et al. reported LDA+{\gwo} gaps of 5.9 and 5.6 eV for \ce{HfO2} and \ce{ZrO2} \cite{Gruning10_PRB}.
Skl\'{e}nard et al. reported PBE+{\gowo} and PBE+{\gwo} gaps of 5.77 and 6.18 eV for \ce{HfO2} \cite{Sklenard18_APL}, about 0.2 eV difference to ours in both but in opposite direction.
This highlights the challenges in arriving at converged $GW$ band gap estimate, which depends sensitively on the starting point and on the computational parameters. $GW$ calculations are highly sensitive to the treatment of core electrons, and show slow convergence with the number of unoccupied states and possibly on basis set, k-points, frequency integration, etc. Recently, Azizi et al. reported that even highly converged {\gowo} calculations on different codes yield 0.1--0.3 eV differences among different implementations \cite{Azizi25_CMS}.
Their best estimate for \ce{ZrO2} using PBEsol+{\gowo} was 5.3--5.4 \cite{Azizi25_CMS}.

In Table \ref{tab:gaps}, we also list experimental results from photoemission spectroscopy (PES) and inverse photoemission spectroscopy (IPS), which should correspond to the fundamental gaps as obtained from $GW$ calculations (i.e., from energies of electron addition and removal), and the results from spectroscopic ellipsometry (SE), which should correspond to the optical gap.
The PBE and r2SCAN band gaps are clearly underestimated, as expected.
Curiously, HSE06 and PBE+\gowo{} results agree within about 0.15 eV for \ce{HfO2} and 0.05 eV for \ce{ZrO2}, HSE06 yielding consistently slightly larger gaps.
HSE06 results are also in slightly better agreement with PES+IPS experiments, 0--0.1 eV for \ce{HfO2} and 0.2--0.4 eV for \ce{ZrO2}.
PBE+\gwo{} gaps, on the other hand, are markedly higher: 0.85 eV and 0.75 eV higher than PBE+{\gowo} gaps for \ce{HfO2} and \ce{ZrO2}, respectively.
Interestingly, the PBE+{\gwo} band gaps are very similar to the HSE+{\gowo} results, within 0.1 eV.

There are several additional factors, however, that affect the comparison to experiments.
First, the above-mentioned problems in converging GW calculations.
Previous benchmarks for $GW$ have resulted in MAE of 0.45 eV for {\gowo} in \cite{Chen15_PRB}, or {\gowo} and {\gwo} MARE of 9.9\% and 5.7\% \cite{Shishkin07_PRB}.
For our materials, this MARE translates to about 0.3--0.4 eV for {\gwo}.
Generally, {\gwo} is expected to be closer to experimental values, although this may depend on the material and on the chosen experimental reference. 

Second, the calculated gaps do not include the finite-temperature effects, i.e., gap renormalization by electron-phonon coupling. Consequently, one would expect the calculated gaps to be higher than experimental values measured at room temperature.
To this end, we evaluated the finite-temperature effects using the one-shot method \cite{Zacharias16_PRB}.
As shown in Fig. S7, this leads to band gaps decreasing by about 0.2 eV and 0.25 eV at 0 K (0.25 eV and 0.3 eV at 300 K) for \ce{HfO2} and \ce{ZrO2}, respectively.
This would recover about half of the discrepancy between our PBE+\gwo{} or HSE06+\gowo{} results and the experiments.

Third, when comparing optical gaps, excitonic effects need to be accounted for.
To the best of our knowledge, the only reported exciton binding energy is
for m-phase \ce{HfO2} calculated by GW+BSE in Ref.\ \cite{Sklenard18_APL}.
They obtained 0.46 eV for m-phase from PBE+QSGW+BSE calculations.
We evaluated the exciton binding energies for c-phase
(due to computational cost), yielding somewhat lower values of 
0.36 eV (dark exciton, 0.34 eV for lowest weak bright exciton) for \ce{HfO2} and 0.32 eV (dark exciton, 0.31 eV for lowest weak bright) for \ce{ZrO2}.
However, taking into account that c-phase dielectric constant is clearly higher than m-phase, the c-phase excitons are likely to be more strongly screened and thus explaining the difference with Ref.\ \cite{Sklenard18_APL}.

Overall, adding these corrections to the HSE+{\gowo} calculations leads to rather good agreement with experiments for \ce{ZrO2}, whereas our values for \ce{HfO2} are higher than experiments by a few tenths of an eV.
Our estimate for {\hzo} is obtained by simple linear extrapolation between the end points.
It is worth noting, that by comparing the experimental and calculated DOS, it was argued in Refs. \cite{Sayan04_JAP,Sklenard18_APL} that PES+IPS numbers suffer from poor extrapolation of band edges and that the fundamental gap could be 6.5--6.7 eV. 

\begin{figure*}[ht]
\begin{center}
  \includegraphics[width=17cm]{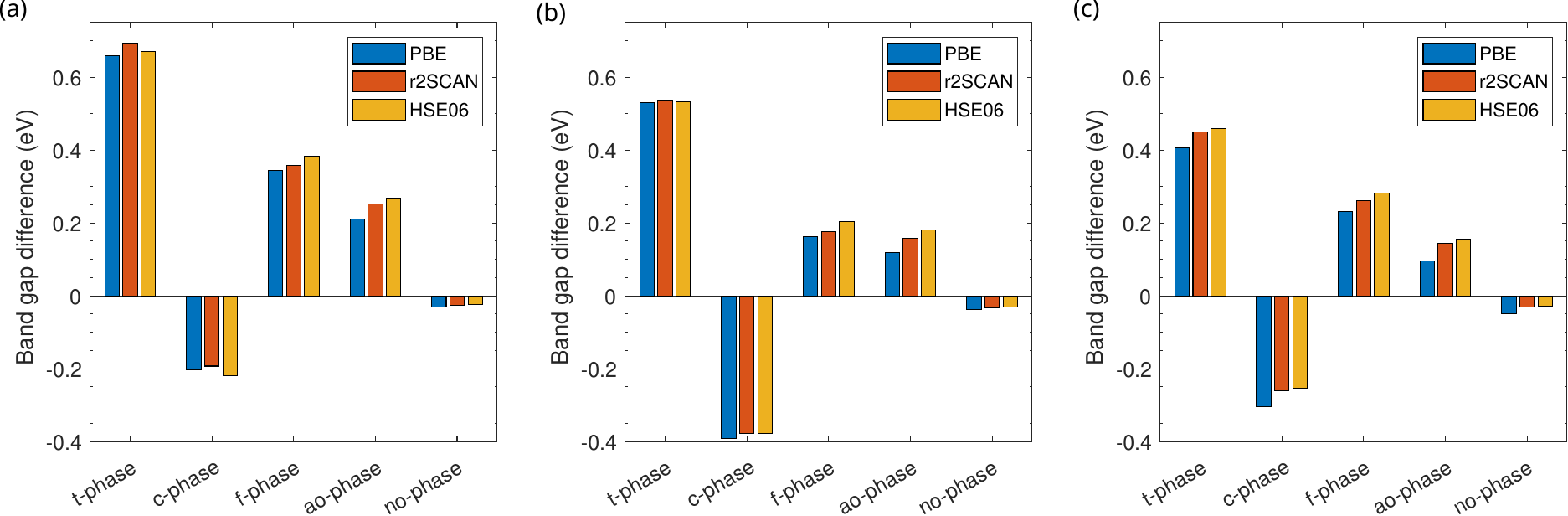}
\end{center}
\caption{\label{fig:allgaps}
The band gap differences with respect to m-phase for (a) \ce{HfO2}, (b) {\hzo}, and (c) \ce{ZrO2}, for all three considered functionals.
}
\end{figure*}

\begin{figure*}[ht]
\begin{center}
  \includegraphics[width=17cm]{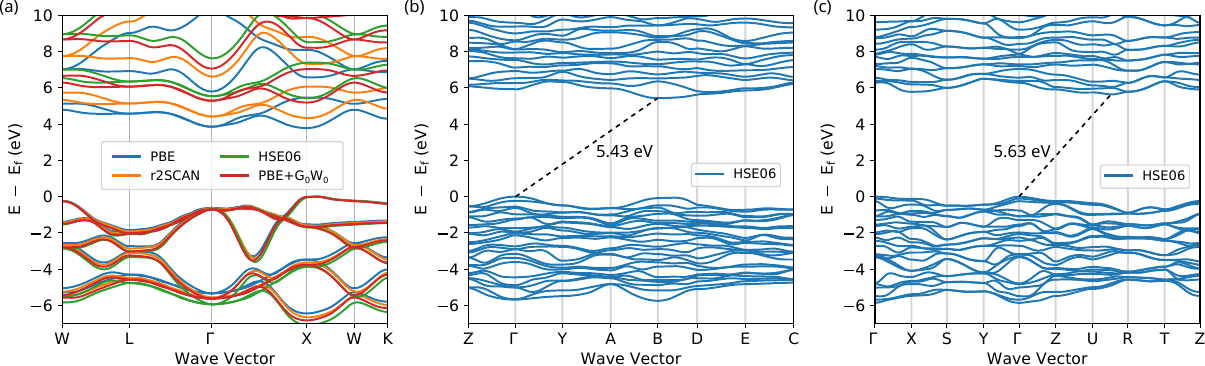}
\end{center}
\caption{\label{fig:bs}
  (a) Band structure of c-phase \ce{HfO2} at different levels of theory. The band structures are aligned at the valence band maximum. 
  (b,c) Band structure of HZO m-phase (b) and f-phase (c) calculated
  with HSE06 functional.
}
\end{figure*}

For other phases, we show the band gap \emph{differences} in Fig.\ \ref{fig:allgaps} (with numerical values listed in Tables S7--S9).
Clearly, the band gap differences are only weakly affected by the functional.
Moreover, this effect is not only present in the band gap, but affects the whole band structure, as shown in Fig. \ref{fig:bs}(a).
Therefore, in order to evaluate band gaps for other phases, one can combine experimental value (or best estimate calculated value) for m-phase and the calculated band gap differences to other phases.
For instance, if we take the final estimated values for indirect gap of {\hzo} from Table~\ref{tab:gaps}, 5.83 eV, and the HSE06 gap differences from Table S8, we obtain gaps of 
6.36, 5.45, 6.03, 6.01, and 5.80 eV, for t-, c-, f-, ao-, and no-phase, respectively. 
Difference between t- and c-phases reaches about 0.9 eV, which can be significant. This may also explain some of the variance in experimentally measured band gaps.
As the band structure for the technologically relevant phases of HZO have not been reported in the literature we show them here, Figure \ref{fig:bs}(b,c). The band structures for all phases calculated using PBE, r2SCAN, and HSE06 functionals are collected in SI Figs. S4--S6.


\subsection{Device modeling}

Finally, we briefly demonstrate the importance of the material parameters in practical applications by carrying out TCAD simulations for ferroelectric capacitor with and without non-ferroelectric interlayers.

\begin{figure*}[ht]
\begin{center}
\includegraphics[width=17cm]{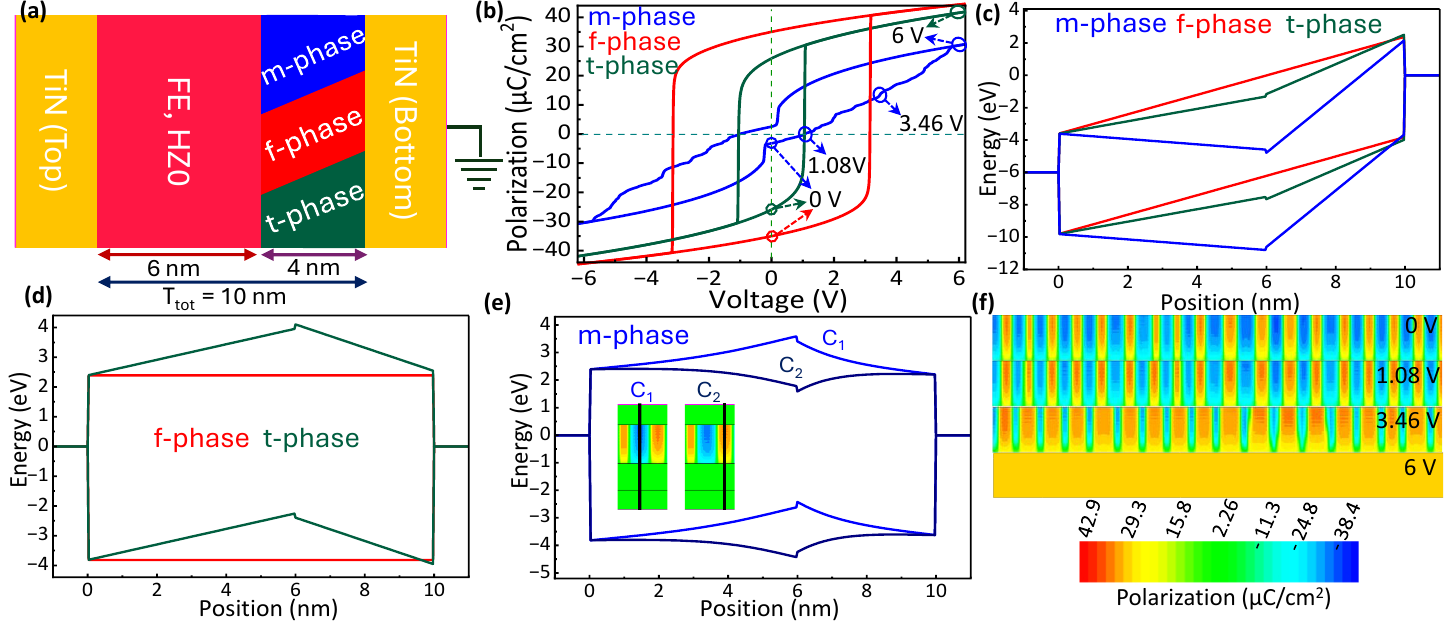}
\end{center}
\caption{\label{fig:tcad}
  (a) Illustration of the device structure used in TCAD simulations. Device with nominal 10 nm FE layer is compared to those with 6 nm FE layer and 4 nm m- or t-phase interlayer. (b) Hysteresis loops from all three cases. (c) Band diagrams from all three cases at the applied bias of 6 V in the saturation regime. (d,e) Band diagrams in the absence of applied bias. In the m-phase, the band diagram through both up- and down-polarized domains (C$_1$ and C$_2$) are shown. (f) Ferroelectric polarization maps over the FE region at different bias points indicated in (b).
}
\end{figure*}

Transmission electron microscopy of Sr and Gd-doped \ce{HfO2} films has revealed that it may contain m-, t-, and f-phases and, particularly in the pristine state, the interface between \ce{HfO2} and the electrode was found to host a region of t-phase \cite{Pesic16_AFM,Grimley16_AEM}.
To this end, we consider how the polarization response of a device changes when the top part of the ferroelectric film is transformed into non-ferroelectric phase, either t- or m-phase, with dramatically different dielectric constants.
The device structure is illustrated in Fig. \ref{fig:tcad}(a). To highlight the effects, we have adopted here a structure with fairly thick non-FE layer. 

We use the rotationally averaged HSE06  dielectric constants and the band gap values evaluated using band gap differences on top of 5.83 eV electronic gap, as listed above. 
For band diagrams, we also need band offsets or electron affinity values. We do not attempt to calculate them here, as explicit interface calculations would be computationally demanding, but also problematic in terms of mismatching lattice systems.
Instead, we use experimentally determined values, although the uncertainties can again be fairly large. In IPS study for \ce{ZrO2}/Si, Sayan et al. found VBO 3.40 and CBO 1.16 eV with measured band gap of 5.68 eV \cite{Sayan04_PSSB}, which translates to an electron affinity $\chi = 4.05-1.16=2.89$ eV.
For \ce{HfO2}/Si, Sayan et al found approx. CBO 1.2 eV (at a gap of 5.7 eV taken from literature) \cite{Sayan02_APL}, translating to $\chi = 2.85$ eV. On the other hand, Afanas'ev et al. found for \ce{HfO2}/Si CBO of 2.0 eV at a measured band gap of 5.6 eV, giving only $\chi=2.05$ eV \cite{Afanasev02_APL}.
{\hzo}/metal valence band offsets were studied using XPS by Jenkins et al. \cite{Jenkins21_ACSAMI}. The average over six metals yielded ionization potential (VBM wrt vacuum) of 8.29 eV. If using our calculated m-phase band gap of 5.83 eV, we arrive at $\chi = 2.46$ eV, which falls in between the two values for \ce{HfO2}.
Clearly, this approach depends sensitively on the band gap value and unfortunately many of these studies do not provide detailed analysis of the phase contents.
Jenkins et al. see weak polarization response, but conclude that the films are not predominately ferroelectric, thus here assigned to the most stable m-phase.
Nevertheless, combining the above results, we will here adopt $\chi_m = 2.46$ eV for m-phase HZO and split the band gap difference between phases equally to valence and conduction band offsets: $\chi_t = 2.195$ eV and $\chi_f = 2.36$ eV for t- and f-phase, respectively. Default value of 4.66 eV is taken for the TiN work function.


In Fig. \ref{fig:tcad}(b) we show how the t- and m-phase non-FE layers affect the hysteresis curves. Fig. \ref{fig:tcad}(c-e) show the corresponding band diagrams from different points of the hysteresis curve. 
The large qualitative difference arises from different depolarization fields. Polarization discontinuity at the FE/non-FE interface results in interface charge that is screened weakly by the m-phase but fairly strongly by the t-phase.
Even with the t-phase there is still non-negligible depolarization field, as evident from Fig. \ref{fig:tcad}(d).
It will be impossible to isolate the properties of the two layers (FE response and layer thicknesses) from this hysteresis curve, without any additional information.

With m-phase the depolarization field is so large that the field inside FE layer exceeds coercive field even in the absence of external bias. 
The small dielectric constant of m-phase means that the interface charge strongly interacts with itself. To reduce this electrostatic energy cost, the system can form domain structure, Fig. \ref{fig:tcad}(f), where the energy is balanced by the polarization gradient term in the GLK formulation. At zero bias, the up- and down-polarized domains have almost the same size. With increasing bias, the domain size of the favored polarization direction increases and eventually the domains disappear, resulting in a homogeneously polarized film.
In the end, the system shows a strongly pinched hysteresis loop, which is sometimes found in pristine devices prior to wake-up \cite{Florent17_TED,Lomenzo19,Siannas22_CP}, and its disappearance during wake-up ascribed to screening of the depolarization field by defects \cite{Siannas22_CP}.
The band diagram shows that this leads to a complex band profile with both electron and hole trap regions at the FE/non-FE interface.


\section{Conclusions}

We carried out systematic first-principles calculations of the structural and electrical properties of \ce{HfO2}, {\hzohalf}, and \ce{ZrO2}, in six polymorphs.
In addition, we compared the results across different levels of theory, from semilocal GGA (PBE) to meta-GGA (r2SCAN) to hybrid functionals (HSE06), and selected calculations at a $GW$-level. The results are always compared to the available experimental results in the literature.

We find that PBE performs poorly for lattice parameters and ionic contribution to dielectric constants, and we thus advise against using PBE for predictions on such properties. On the other hand, r2SCAN and HSE06 were found to yield generally good lattice parameters and response properties.
The energy differences between different phases are similar with all functionals. r2SCAN yields somewhat lower energy for the f- and ao-phases, but not enough to lead to qualitative differences in their stability.

The band gaps calculated at different levels of theory differ dramatically, as expected. However, the band gap differences among different phases are
very similar. 
Therefore, although there is still uncertainty in the absolute band gap values, one can use the band gap differences as long as experimental band gap is known for one phase.
By adding finite-temperature band-gap renormalization to the $GW$ results (and excitonic binding energy in the case of optical gaps), a reasonable agreement with experiments could be obtained.

The motivation for our study was to extract material parameters needed in TCAD simulations of ferroelectric devices. To this end, we carried out TCAD simulations for {\hzo} FeCAP, and showed how the phase content of an unintentional interlayer can affect the device characteristics. Due to the large difference in the dielectric constant of m- and t-phases, the device may switch from conventional ferroelectric hysteresis to pinched hysteresis after the onset of domain formation.

\bibliography{refs.bib}

\end{document}


\title{Supporting Information for "Structural and electronic properties of (Hf,Zr)O$_2$ phases from first principles"}
\author{Sibghatullah Khan}
\author{Hannu-Pekka Komsa}
\email{hannu-pekka.komsa@oulu.fi}
\affiliation{%
Advanced Electronics Centre, Faculty of Information Technology and Electrical Engineering, University of Oulu, P.O. Box 4500, Oulu, FIN-90014, Finland
}%

\maketitle


\section{$GW$ parameters}

\begin{figure*}[!h]
\begin{center}
  \includegraphics[width=\textwidth]{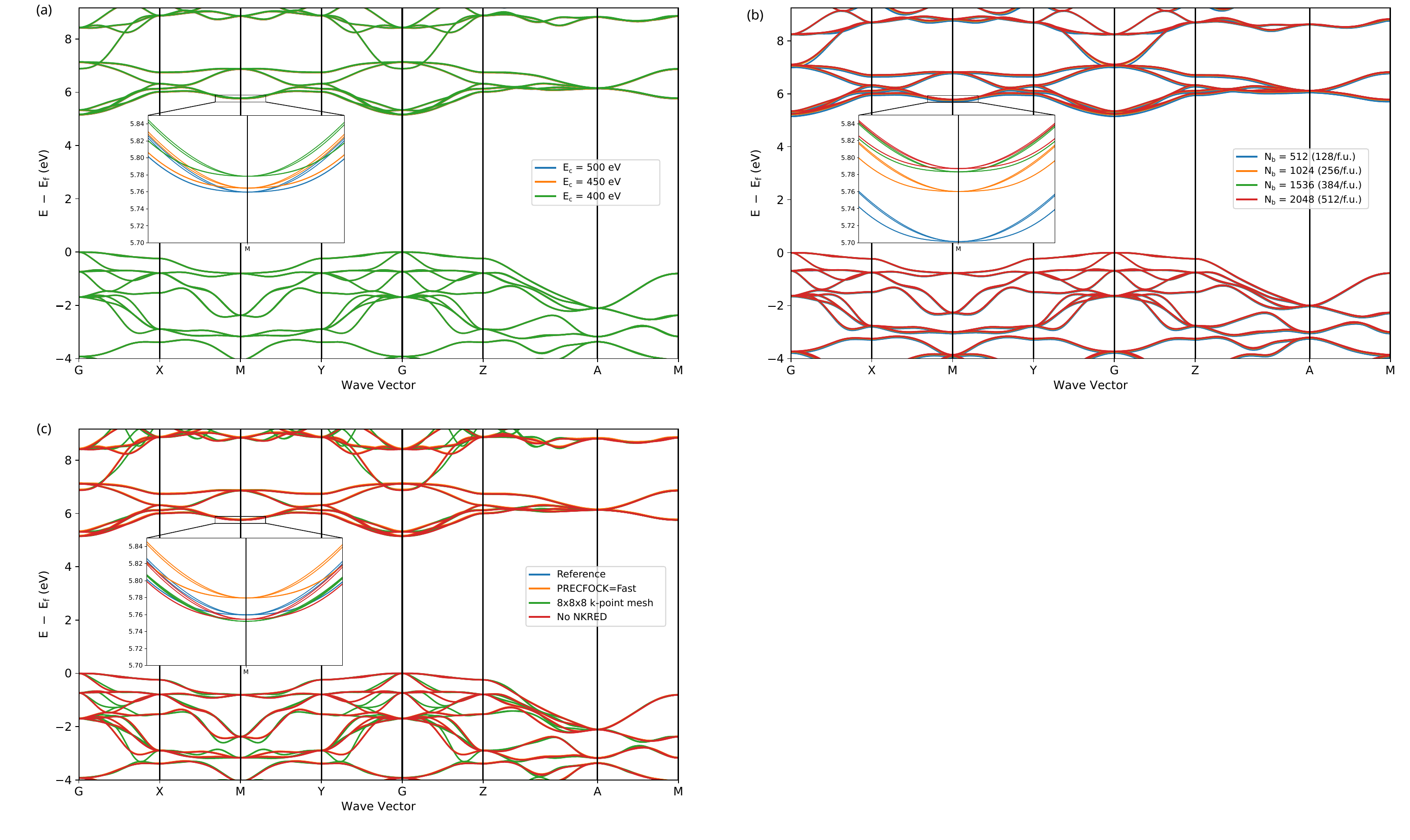}
\end{center}
\caption{\label{fig:gwtests}
  Band structures from $GW$ parameter tests carried out on
  a conventional unit cell of c-phase \ce{HfO2} (4 formula units).
  (a) The effect of plane-wave energy cutoff.
  (b) The effect of number of bands in the calculation.
  (c) The effect of k-point sampling and Fock exchange potential approximations. Reference has 6x6x6 k-point mesh, downsampled k-point mesh in Fock operator (NKRED=2) and standard FFT grid (PRECFOCK=Normal).
}
\end{figure*}

Benchmarking results for $GW$ calculations are collected in
Fig.~\ref{fig:gwtests}.
The band structures are aligned at valence band maximum located
at the $\Gamma$-point and the inset shows zoom-in on the conduction
band valley minimum at M-point. 

\section{Lattice parameters}

Tables \ref{tab:HfO2}, \ref{tab:HZO}, and \ref{tab:ZrO2}
give energy with respect to the monoclinic-phase (in eV/formula unit),
lattice constants (in {\AA}, half-size in orthorhombic phases),
and cell volume (in {\AA}$^3$, half-size in orthorhombic phases)
from DFT calculations with all three functionals used in this work.
These are also compared to experimental values and selected
previous computational results, with the relevant reference given.
The last r2SCAN results are from Materials Project database,
in which case the material-ID is given \cite{mp}.




\begin{table}[!ht]
\begin{center}
\caption{\label{tab:HfO2}
  Lattice parameters and energy differences of HfO$_2$ phases.
}
\footnotesize
\begin{tabular}{lcccccc}
          & $\Delta E$ & a & b & c & V & Ref. \\
\hline
\multicolumn{7}{c}{m-phase} \\
PBE    & 0.000 & 5.147 & 5.199 & 5.329 & 140.56 \\
r2SCAN & 0.000 & 5.100 & 5.167 & 5.273 & 137.03 \\
HSE06  & 0.000 & 5.104 & 5.159 & 5.272 & 136.90 \\
expt      &       & 5.12  & 5.17  & 5.29  & 137.9 & \cite{Ruh70} \\  
expt      &       & 5.07  & 5.14  & 5.29  & 135.8 & \cite{Muller12_NL} \\  
LDA       &       & 5.11  & 5.16  & 5.28  & 137.1 & \cite{Materlik15_JAP} \\
GGA       & 0.0   & 5.13  & 5.19  & 5.30  & 139.0 & \cite{Jaffe05_PRB} \\  
r2SCAN    & 0.0   & 5.09  & 5.15  & 5.28  & 136.52 & mp-352 \\  
\hline
\multicolumn{7}{c}{t-phase} \\
PBE    & 0.165 & 5.082 & 5.082 & 5.229 & 135.06 \\
r2SCAN & 0.150 & 5.042 & 5.042 & 5.165 & 131.32 \\
HSE06  & 0.157 & 5.038 & 5.038 & 5.155 & 130.83 \\
expt      &       & 5.06  & 5.06  & 5.20  & 133.1 & \cite{Ruh70,MacLaren09} \\  
LDA       & 0.092 & 5.03  & 5.03  & 5.12  & 129.5 & \cite{Materlik15_JAP} \\
GGA       & 0.156 & 5.06  & 5.06  & 5.20  & 133.1 & \cite{Jaffe05_PRB} \\  
r2SCAN    & 0.153 & 5.035 & 5.035 & 5.17  & 130.94 & mp-1018721 \\  
\hline
\multicolumn{7}{c}{c-phase} \\
PBE    & 0.261 & 5.079 & 5.079 & 5.079 & 131.05 \\ 
r2SCAN & 0.234 & 5.040 & 5.040 & 5.040 & 127.99 \\
HSE06  & 0.230 & 5.036 & 5.036 & 5.036 & 127.71 \\
expt      &       & 5.08  & 5.08  & 5.08  & 131.1 & \cite{ElShanshoury70} \\  
LDA       & 0.137 & 5.03  & 5.03  & 5.03  & 127.3 & \cite{Materlik15_JAP} \\
GGA       & 0.237 & 5.06  & 5.06  & 5.06  & 129.6 & \cite{Jaffe05_PRB} \\  
r2SCAN    & 0.252 & 5.03  & 5.03  & 5.03  & 127.43 & mp-550893 \\  
\hline
\multicolumn{7}{c}{f-phase} \\
PBE    & 0.083 & 5.056 & 5.274 & 5.086 & 135.59 \\
r2SCAN & 0.065 & 5.016 & 5.226 & 5.043 & 132.20 \\
HSE06  & 0.076 & 5.013 & 5.221 & 5.040 & 131.92 \\
expt & & & & & \\
LDA       & 0.062 & 5.02  & 5.22  & 5.04  & 132.1 & \cite{Materlik15_JAP} \\
GGA       & >0.03 & 5.01  & 5.29  & 5.08  & 134.6 & \cite{Huan14_PRB}\\ 
r2SCAN    & 0.066 & 5.00  & 5.22  & 5.03  & 131.46 & mp-685097 \\  
\hline
\multicolumn{7}{c}{ao-phase} \\
PBE    & 0.071 & 5.033 & 5.252 & 5.089 & 134.55 \\
r2SCAN & 0.048 & 4.992 & 5.213 & 5.046 & 131.30 \\
HSE06  & 0.061 & 4.993 & 5.206 & 5.043 & 131.08 \\
expt      &        & 5.01  & 5.23  & 5.06  & 132.6 & \cite{Suyama85} \\  
LDA       & 0.024  & 5.07  & 5.29  & 5.15  & 138.4 & \cite{Materlik15_JAP} \\
GGA       & 0.065  & 4.92  & 5.16  & 4.96  & 134.1 & \cite{Jaffe05_PRB} \\  
r2SCAN    & 0.048  & 4.985 & 5.20  & 5.04  & 130.68 & mp-1858 \\  
\hline
\multicolumn{7}{c}{no-phase} \\
PBE    & 0.027 & 5.181 & 5.323 & 5.099 & 140.60 \\
r2SCAN & 0.029 & 5.146 & 5.274 & 5.057 & 137.23 \\
HSE06  & 0.028 & 5.141 & 5.270 & 5.060 & 137.10 \\
r2SCAN     & 0.030 & 5.14 & 5.26  & 5.045  & 136.46 & mp-775757 \\  
\end{tabular}
\end{center}
\end{table}

\begin{table}[!ht]
\begin{center}
\caption{\label{tab:HZO}
  Lattice parameters and energy differences of {\hzo} phases.
}
\footnotesize
\begin{tabular}{lcccccc}
          & $\Delta E$ & a & b & c & V & Ref. \\
\hline
\multicolumn{7}{c}{m-phase} \\
PBE    & 0.000 & 5.168 & 5.224 & 5.351 & 142.42 \\  
r2SCAN & 0.000 & 5.128 & 5.199 & 5.301 & 139.40 \\
HSE06  & 0.000 & 5.128 & 5.184 & 5.295 & 138.81 \\
LDA       & 0.0   & 5.11  & 5.18  & 5.28  & 137.6 & \cite{Materlik15_JAP} \\  
\hline
\multicolumn{7}{c}{t-phase} \\
PBE    & 0.137 & 5.102 & 5.103 & 5.254 & 136.80 \\  
r2SCAN & 0.115 & 5.069 & 5.070 & 5.195 & 133.51 \\
HSE06  & 0.127 & 5.058 & 5.059 & 5.177 & 132.50 \\
LDA    & 0.070 & 5.04  & 5.04  & 5.13  & 130.3 & \cite{Materlik15_JAP} \\  
\hline
\multicolumn{7}{c}{c-phase} \\
PBE    & 0.237 & 5.100 & 5.100 & 5.100 & 132.65 \\
r2SCAN & 0.201 & 5.066 & 5.065 & 5.065 & 129.97 \\
HSE06  & 0.201 & 5.054 & 5.054 & 5.054 & 129.10 \\
LDA    & 0.119 & 5.03  & 5.03  & 5.03  & 127.3 & \cite{Materlik15_JAP} \\  
\hline
\multicolumn{7}{c}{f-phase} \\
PBE    & 0.077 & 5.078 & 5.298 & 5.108 & 137.42 \\  
r2SCAN & 0.056 & 5.046 & 5.255 & 5.072 & 134.50 \\
HSE06  & 0.069 & 5.038 & 5.245 & 5.064 & 133.82 \\
expt   & -     & 5.01  & 5.24  & 5.04  & 132.3 & \cite{Muller12_NL} \\  
LDA    & 0.049 & 5.03  & 5.23  & 5.05  & 132.8 & \cite{Materlik15_JAP} \\ 
\hline
\multicolumn{7}{c}{ao-phase} \\
PBE    & 0.068 & 5.055 & 5.275 & 5.111 & 136.27 \\  
r2SCAN & 0.043 & 5.021 & 5.242 & 5.074 & 133.52 \\
HSE06  & 0.058 & 5.015 & 5.229 & 5.063 & 132.77 \\
LDA    & 0.025 & 5.06  & 5.27  & 5.14  & 137.1 & \cite{Materlik15_JAP} \\ 
\hline
\multicolumn{7}{c}{no-phase} \\
PBE    & 0.027 & 5.206 & 5.342 & 5.120 & 142.38 \\  
r2SCAN & 0.028 & 5.180 & 5.298 & 5.084 & 139.52 \\
HSE06  & 0.027 & 5.161 & 5.296 & 5.084 & 138.97 \\
\end{tabular}
\end{center}
\end{table}

\begin{table}[!ht]
\begin{center}
\caption{\label{tab:ZrO2}
  Lattice parameters and energy differences of ZrO$_2$ phases.
}
\footnotesize
\begin{tabular}{lcccccc}
          & $\Delta E$ & a & b & c & V & Ref. \\
\hline
\multicolumn{7}{c}{m-phase} \\
PBE    & 0.000 & 5.190 & 5.247 & 5.375 & 144.29 \\
r2SCAN & 0.000 & 5.156 & 5.229 & 5.332 & 141.83 \\
HSE06  & 0.000 & 5.148 & 5.209 & 5.314 & 140.53 \\
expt.     &       & 5.15  & 5.21  & 5.32  & 140.88 & \cite{Howard88,Kisi98}\\ 
LDA       & 0.0   & 5.11  & 5.20  & 5.28  & 138.2 & \cite{Materlik15_JAP} \\
GGA       & 0.0   & 5.20  & 5.28  & 5.35  & 144.7 &\cite{Jaffe05_PRB} \\  
GGA       & 0.0   & 5.19  & 5.24  & 5.38  & 144.1 & \cite{Fadda10_PRB}\\  
r2SCAN    & 0.0   & 5.15  & 5.23  & 5.33  & 141.67 & mp-2858 \\  
\hline
\multicolumn{7}{c}{t-phase} \\
PBE    & 0.111 & 5.123 & 5.123 & 5.278 & 138.48 \\
r2SCAN & 0.081 & 5.096 & 5.096 & 5.225 & 135.67 \\
HSE06  & 0.098 & 5.078 & 5.078 & 5.196 & 133.96 \\
expt      &       & 5.11 & 5.11 & 5.212 & 136.00 & \cite{Igawa01} \\  
expt      &       & 5.09 & 5.09 & 5.18 & 134.2 & \cite{Kisi98} \\  
expt      &       & 5.08 & 5.08 & 5.17 & 133.4 & \cite{Muller12_NL} \\  
LDA       & 0.049 & 5.05  & 5.05  & 5.15  & 131.3 & \cite{Materlik15_JAP} \\
GGA       & 0.109 & 5.13  & 5.13  & 5.21  & 137.9 & \cite{Jaffe05_PRB} \\  
GGA       & 0.112 & 5.12  & 5.12  & 5.28  & 138.4 & \cite{Fadda10_PRB} \\  
r2SCAN    & 0.081 & 5.09  & 5.09  & 5.23  & 135.44 & mp-2574 \\  
\hline
\multicolumn{7}{c}{c-phase} \\
PBE    & 0.212 & 5.118 & 5.118 & 5.118 & 134.02 \\
r2SCAN & 0.167 & 5.092 & 5.092 & 5.092 & 132.00 \\
HSE06  & 0.170 & 5.073 & 5.073 & 5.073 & 130.58 \\
expt      &       & 5.09  & 5.09  & 5.09  & 131.87 & \cite{Katz71} \\  
expt      &       & 5.12  & 5.12  & 5.12  & 134.2 & \cite{} \\  
LDA       & 0.120 & 5.03  & 5.03  & 5.03  & 127.3 & \cite{Materlik15_JAP} \\
GGA       & 0.171 & 5.13  & 5.13  & 5.13  & 134.8 & \cite{Jaffe05_PRB} \\  
GGA       & 0.215 & 5.12  & 5.12  & 5.12  & 134.9 & \cite{Fadda10_PRB} \\  
r2SCAN    & 0.168 & 5.09  & 5.09  & 5.09  & 131.88 & mp-1565 \\  
\hline
\multicolumn{7}{c}{f-phase} \\
PBE    & 0.072 & 5.101 & 5.323 & 5.129 & 139.26 \\
r2SCAN & 0.050 & 5.076 & 5.287 & 5.100 & 136.87 \\
HSE06  & 0.065 & 5.062 & 5.268 & 5.085 & 135.58 \\
expt      &       & 5.068 & 5.260 & 5.077 & 135.34 & \cite{Kisi89} \\  
LDA       & 0.037 & 5.04  & 5.24  & 5.05  & 133.4 & \cite{Materlik15_JAP} \\
r2SCAN    & 0.051 & 5.07  & 5.28  & 5.10  & 136.77 & m-556605 \\  
\hline
\multicolumn{7}{c}{ao-phase} \\
PBE    & 0.067 & 5.076 & 5.299 & 5.132 & 138.02 \\
r2SCAN & 0.040 & 5.049 & 5.272 & 5.101 & 135.78 \\
HSE06  & 0.057 & 5.038 & 5.251 & 5.085 & 134.52 \\
expt      &       & 5.04  & 5.26  & 5.09  & 134.9 & \cite{Ohtaka90} \\  
LDA       & 0.025 & 5.07  & 5.27  & 5.20  & 138.7 & \cite{Materlik15_JAP} \\
GGA       & 0.049 & 5.09  & 5.31  & 5.14  & 138.8 & \cite{Jaffe05_PRB} \\  
GGA       & 0.067 & 5.08  & 5.30  & 5.13  & 138.1 & \cite{Fadda10_PRB} \\  
r2SCAN    & 0.039 & 5.045 & 5.27  & 5.10  & 135.63 & mp-1190186 \\  
\hline
\multicolumn{7}{c}{no-phase} \\
PBE    & 0.026 & 5.230 & 5.363 & 5.141 & 144.20 \\
r2SCAN & 0.028 & 5.211 & 5.325 & 5.113 & 141.88 \\
HSE6   & 0.027 & 5.187 & 5.315 & 5.107 & 140.79 \\
r2SCAN    & 0.027 & 5.22 & 5.31  & 5.11  & 141.57 & mp-776404 \\  
\end{tabular}
\end{center}
\end{table}

\clearpage

\section{Dielectric constants}


Numerical values for calculated dielectric constants are collected
in Tables \ref{tab:dielHO}, \ref{tab:dielHZO} and \ref{tab:dielZO},
for all phases and functionals.
Both ionic and electronics contributions are given, as the first
and second values, respectively.

\begin{table}[!ht]
\begin{center}
\caption{\label{tab:dielHO}
  Dielectric constants of HfO$_2$ phases.
}
\begin{tabular}{lccccc}
          & xx & yy & zz & avg \\
\hline
\multicolumn{5}{c}{m-phase} \\
PBE    & 15.5+4.9 & 13.9+4.8 & 11.1+4.6 & 13.5+4.8 \\
r2SCAN & 13.8+4.5 & 12.5+4.5 & 10.0+4.3 & 12.1+4.5 \\
HSE06  & 13.6+4.3 & 12.3+4.3 & 9.9+4.1 & 12.0+4.2 \\
\multicolumn{5}{c}{t-phase} \\
PBE    & 207.2+5.2 & 207.2+5.2 & 12.7+4.7 & 142.4+5.0 \\
r2SCAN & 92.0+4.8 & 92.0+4.8 & 12.2+4.4 & 65.4+4.7 \\
HSE06  & 75.5+4.5 & 75.5+4.5 & 12.5+4.2 & 54.5+4.4 \\
\multicolumn{5}{c}{c-phase} \\
PBE    & 31.6+5.2 & 31.6+5.2 & 31.6+5.2 & 31.6+5.2 \\
r2SCAN & 26.3+4.8 & 26.3+4.8 & 26.3+4.8 & 26.3+4.8 \\
HSE06  & 24.6+4.5 & 24.6+4.5 & 24.6+4.5 & 24.6+4.5 \\
\multicolumn{5}{c}{f-phase} \\
PBE    & 19.8+5.1 & 15.2+4.9 & 18.4+4.9 & 17.8+5.0 \\
r2SCAN & 17.1+4.8 & 13.5+4.6 & 16.3+4.6 & 15.6+4.6 \\
HSE06  & 16.8+4.5 & 13.3+4.3 & 16.3+4.3 & 15.4+4.4 \\
\multicolumn{5}{c}{ao-phase} \\
PBE    & 19.5+5.1 & 15.7+5.0 & 14.9+4.9 & 16.7+5.0 \\
r2SCAN & 17.0+4.8 & 13.9+4.7 & 13.5+4.6 & 14.8+4.7 \\
HSE06  & 0.0+0.0 & 0.0+0.0 & 0.0+0.0 & 0.0+0.0 \\
\multicolumn{5}{c}{no-phase} \\
PBE    & 14.1+4.8 & 11.6+4.6 & 16.4+4.9 & 14.0+4.8 \\
r2SCAN & 12.6+4.5 & 10.3+4.3 & 14.6+4.6 & 12.5+4.5 \\
HSE06  & 0.0+0.0 & 0.0+0.0 & 0.0+0.0 & 0.0+0.0 \\
\end{tabular}
\end{center}
\end{table}

\begin{table}[!ht]
\begin{center}
\caption{\label{tab:dielHZO}
  Dielectric constants of {\hzo} phases.
}
\begin{tabular}{lccccc}
          & xx & yy & zz & avg \\
\hline
\multicolumn{5}{c}{m-phase} \\
PBE    & 17.1+5.1 & 15.3+5.1 & 12.1+4.8 & 14.8+5.0 \\
r2SCAN & 15.4+4.7 & 14.0+4.8 & 10.9+4.5 & 13.4+4.7 \\
HSE06  & 15.0+4.5 & 13.6+4.4 & 11.0+4.2 & 13.2+4.4 \\
LDA       & 21.5+5.2 & 19.6+5.2 & 15.6+4.9 & 18.9+5.2 \\  
\multicolumn{5}{c}{t-phase} \\
PBE    & 181.1+5.4 & 152.1+5.4 & 13.9+4.8 & 115.7+5.2 \\
r2SCAN & 91.6+5.0 & 79.7+5.0 & 13.4+4.6 & 61.5+4.9 \\
HSE06  & 73.5+4.7 & 68.1+4.7 & 13.7+4.3 & 51.8+4.6 \\
LDA       & 54.3+6.1 & 54.3+6.1 & 19.9+5.4 & 42.8+5.9 \\  
\multicolumn{5}{c}{c-phase} \\
PBE    & 36.3+5.5 & 36.6+5.5 & 36.6+5.5 & 36.5+5.5 \\
r2SCAN & 29.9+5.1 & 30.0+5.1 & 30.0+5.1 & 30.0+5.1 \\
HSE06  & 27.1+4.7 & 27.2+4.7 & 27.2+4.7 & 27.2+4.7 \\
\multicolumn{5}{c}{f-phase} \\
PBE    & 22.7+5.4 & 16.8+5.1 & 21.1+5.1 & 20.2+5.2 \\
r2SCAN & 19.6+5.0 & 15.0+4.7 & 18.8+4.8 & 17.8+4.8 \\
HSE06  & 19.0+4.7 & 14.8+4.5 & 18.6+4.5 & 17.5+4.5 \\
LDA       & 26.0+5.5 & 20.7+5.2 & 24.6+5.3 & 23.8+5.3 \\  
\multicolumn{5}{c}{ao-phase} \\
PBE    & 22.5+5.4 & 17.7+5.2 & 16.6+5.1 & 18.9+5.3 \\
r2SCAN & 19.4+5.0 & 15.6+4.9 & 15.1+4.7 & 16.7+4.9 \\
HSE06  & 0.0+0.0 & 0.0+0.0 & 0.0+0.0 & 0.0+0.0 \\
\multicolumn{5}{c}{no-phase} \\
PBE    & 15.8+5.1 & 12.8+4.8 & 18.3+5.2 & 15.6+5.0 \\
r2SCAN & 14.1+4.7 & 11.3+4.5 & 16.2+4.8 & 13.9+4.7 \\
HSE06  & 0.0+0.0 & 0.0+0.0 & 0.0+0.0 & 0.0+0.0 \\
\end{tabular}
\end{center}
\end{table}

\begin{table}[!ht]
\begin{center}
\caption{\label{tab:dielZO}
  Dielectric constants of ZrO$_2$ phases.
}
\begin{tabular}{lccccc}
          & xx & yy & zz & avg \\
\hline
\multicolumn{5}{c}{m-phase} \\
PBE    & 18.9+5.3 & 17.2+5.4 & 13.2+5.0 & 16.4+5.2 \\
r2SCAN & 17.2+4.9 & 15.6+5.0 & 12.0+4.7 & 14.9+4.9 \\
HSE06  & 16.6+4.6 & 15.1+4.6 & 11.9+4.4 & 14.5+4.5 \\
\hline
\multicolumn{5}{c}{t-phase} \\
PBE    & 109.0+5.7 & 109.0+5.7 & 13.4+5.0 & 77.1+5.5 \\
r2SCAN & 65.6+5.3 & 65.6+5.3 & 13.1+4.7 & 48.1+5.1 \\
HSE06  & 58.0+4.9 & 58.0+4.9 & 14.0+4.5 & 43.3+4.8 \\
\hline
\multicolumn{5}{c}{c-phase} \\
PBE    & 41.6+5.8 & 41.6+5.8 & 41.6+5.8 & 41.6+5.8 \\
r2SCAN & 34.2+5.3 & 34.2+5.3 & 34.2+5.3 & 34.2+5.3 \\
HSE06  & 30.2+4.9 & 30.2+4.9 & 30.2+4.9 & 30.2+4.9 \\
\hline
\multicolumn{5}{c}{f-phase} \\
PBE    & 25.8+5.7 & 18.4+5.3 & 24.3+5.4 & 22.8+5.4 \\
r2SCAN & 23.7+5.6 & 17.2+5.3 & 23.1+5.4 & 21.4+5.4 \\
HSE06  & 21.5+4.9 & 16.2+4.6 & 21.5+4.7 & 19.7+4.7 \\
\hline
\multicolumn{5}{c}{ao-phase} \\
PBE    & 25.7+5.7 & 19.9+5.5 & 18.7+5.3 & 21.4+5.5 \\
r2SCAN & 22.1+5.2 & 17.5+5.1 & 17.0+4.9 & 18.9+5.1 \\
HSE06  & 0.0+0.0 & 0.0+0.0 & 0.0+0.0 & 0.0+0.0 \\
\hline
\multicolumn{5}{c}{no-phase} \\
PBE    & 17.8+5.3 & 14.1+5.1 & 20.2+5.4 & 17.4+5.3 \\
r2SCAN & 15.9+4.9 & 12.5+4.7 & 18.0+5.0 & 15.5+4.9 \\
HSE06  & 0.0+0.0 & 0.0+0.0 & 0.0+0.0 & 0.0+0.0 \\
\end{tabular}
\end{center}
\end{table}

\clearpage

\section{Infrared spectra}

The infrared spectra are calculated from Born effective charges and the vibrational normal modes, which are obtained from the dielectric constant calculations.
Non-analytical corrections to phonon frequencies are not applied herein.
The spectra are broadened using a Lorentzian line shape, defined by half-width-half-maximum scale parameter $\gamma$.

\begin{figure*}[ht]
\begin{center}
  \includegraphics[width=10cm]{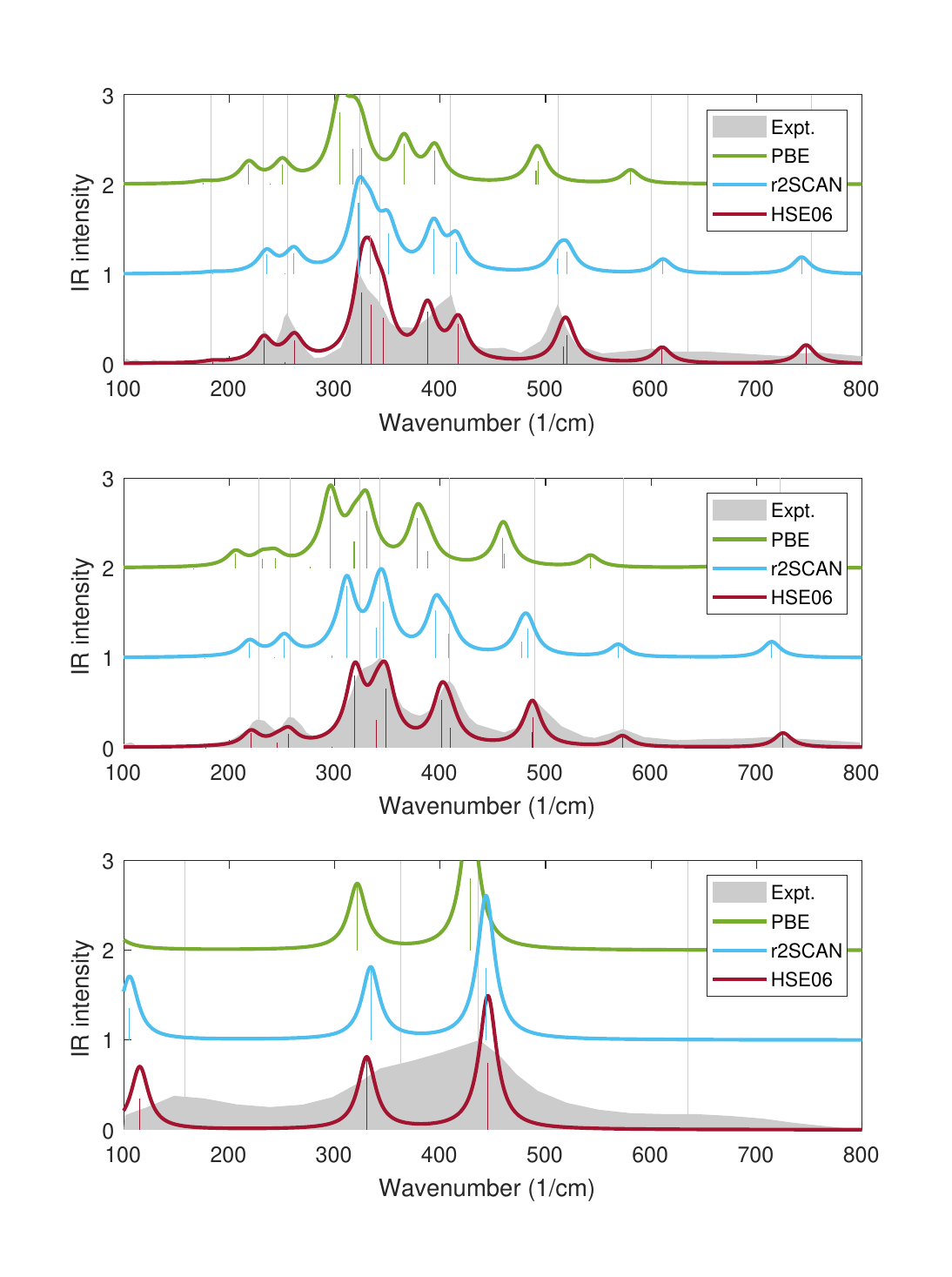}
\end{center}
\caption{\label{fig:irxc}
  Simulated infrared spectra from all three functionals for the three phases with available experimental data: \ce{HfO2} m-phase, \ce{ZrO2} m-phase, and \ce{ZrO2} t-phase.
  The vertical line segments indicate the positions and intensities of the IR-active modes.
  The combined spectra are broadened using $\gamma = 20$ cm$^{-1}$ to allow comparison to experiments, shown with filled light gray spectra where available.
}
\end{figure*}

Experimental data is available for the \ce{HfO2} m-phase and \ce{ZrO2} m- and t-phases \cite{Neumayer01_JAP}.
Comparison of the experimental data with all three functionals are shown in Fig.~\ref{fig:irxc}. 
r2SCAN and HSE yield good agreement for all three cases (and with each other). The only clear discrepancy is found for the 150 cm$^{-1}$ mode of t-phase \ce{ZrO2}.
PBE underestimates the mode wave numbers, in line with the underbinding seen with the lattice parameters. The peak intensities, on the other hand, agree quite closely with all three functionals.

\begin{figure*}[ht]
\begin{center}
  \includegraphics[width=16cm]{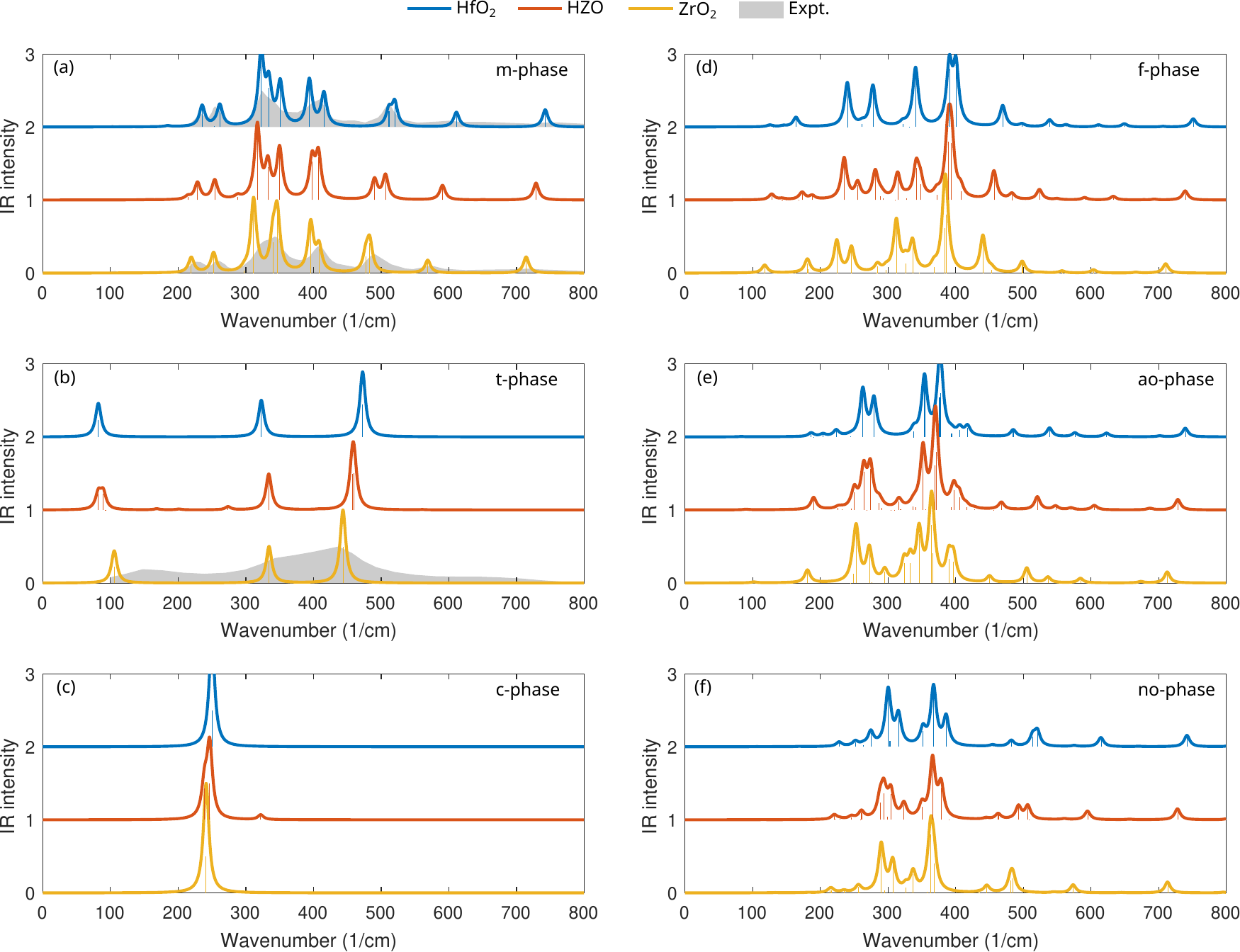}
\end{center}
\caption{\label{fig:irall}
  Simulated infrared spectra of all materials and phases calculated using r2SCAN functional.
  The vertical line segments indicate the positions and intensities of the IR-active modes.
  The combined spectra are broadened using $\gamma = 10$ cm$^{-1}$.
  Experimental spectra are shown with filled light gray spectra where available.
}
\end{figure*}

The IR spectra for all considered phases calculated using r2SCAN are collected in Fig.~\ref{fig:irall}.

Raman, IR, and XRD spectra for 10 phases of \ce{HfO2} and \ce{ZrO2} were simulated using LDA in \cite{Kersch22_FN}.
The results are largely consistent with ours. LDA is found to alse give IR spectrum in close agreement with experiments, and even reproduces the 150 cm$^{-1}$ mode for t-phase \ce{ZrO2}.
This agrees with previous benchmarking by He et al.\cite{He14_PRB}, where LDA was found to yield underestimated lattice constants for some common oxides, yet giving vibrational properties close to experiments (much better than GGAs).

\clearpage

\section{Band structure}

\begin{figure*}[!h]
\begin{center}
  \includegraphics[width=0.5\textwidth]{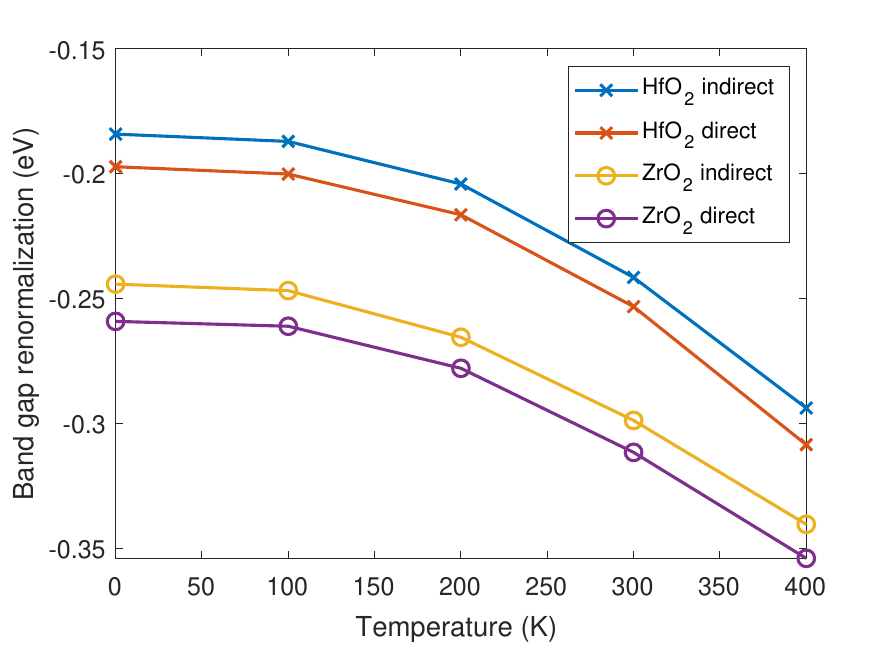}
\end{center}
\caption{\label{fig:gapT}
  Band gap renormalization due to electron-phonon coupling,
  for m-phase \ce{HfO2} and \ce{ZrO2} evaluated at PBE level.
}
\end{figure*}

Numerical values of the band gaps are collected in Tables \ref{tab:gapsHO}, \ref{tab:gapsHZO}, and \ref{tab:gapsZO}.
Band structures are shown in Figs.\ \ref{fig:bs_ho1}, \ref{fig:bs_hzo1}, and \ref{fig:bs_zo1}. The k-point directions are the same as in Ref.\ \cite{Jiang10_PRB}.
Finally, the band gap renormalizations are shown in
Fig.\ \ref{fig:gapT}.


\begin{table}[!ht]
\begin{center}
\caption{\label{tab:gapsHO}
  Band gaps of HfO$_2$ phases.
}
\begin{tabular}{lccc}
          & indirect & direct & diff. \\
\hline
\multicolumn{4}{c}{m-phase} \\
PBE    & 3.998 & 4.083 & 0.000 \\
r2SCAN & 4.545 & 4.630 & 0.000 \\
HSE06  & 5.670 & 5.762 & 0.000 \\
PBE+\gowo & 5.518 & 5.609 & 0.000 \\
HSE06+\gowo & 6.267 & 6.366 & 0.000 \\
\multicolumn{4}{c}{t-phase} \\
PBE    & 4.656 & 4.731 & 0.658 \\
r2SCAN & 5.238 & 5.321 & 0.693 \\
HSE06  & 6.341 & 6.429 & 0.671 \\
PBE+\gowo & 6.179 & 6.264 & 0.661 \\
HSE06+\gowo & 6.900 & 6.998 & 0.633 \\
\multicolumn{4}{c}{c-phase} \\
PBE    & 3.794 & 3.794 & -0.203 \\
r2SCAN & 4.352 & 4.352 & -0.193 \\
HSE06  & 5.450 & 5.450 & -0.220 \\
PBE+\gowo & 5.206 & 5.206 & -0.312 \\
HSE06+\gowo & 5.921 & 5.921 & -0.346 \\
\multicolumn{4}{c}{f-phase} \\
PBE    & 4.342 & 4.593 & 0.345 \\
r2SCAN & 4.904 & 5.137 & 0.359 \\
HSE06  & 6.053 & 6.317 & 0.383 \\
\multicolumn{4}{c}{ao-phase} \\
PBE    & 4.208 & 4.409 & 0.210 \\
r2SCAN & 4.797 & 4.987 & 0.252 \\
HSE06  & 5.938 & 6.116 & 0.268 \\
\multicolumn{4}{c}{no-phase} \\
PBE    & 3.967 & 3.967 & -0.031 \\
r2SCAN & 4.519 & 4.519 & -0.026 \\
HSE06  & 5.647 & 5.647 & -0.023 \\
%
\end{tabular}
\end{center}
\end{table}

\begin{table}[!ht]
\begin{center}
\caption{\label{tab:gapsHZO}
  Band gaps of {\hzo} phases.
}
\begin{tabular}{lccc}
          & indirect & direct & diff. \\
\hline
\multicolumn{4}{c}{m-phase} \\
PBE    & 3.738 & 3.817 & 0.000 \\
r2SCAN & 4.272 & 4.351 & 0.000 \\
HSE06  & 5.422 & 5.506 & 0.000 \\
\multicolumn{4}{c}{t-phase} \\
PBE    & 4.269 & 4.344 & 0.531 \\
r2SCAN & 4.809 & 4.894 & 0.538 \\
HSE06  & 5.955 & 6.045 & 0.533 \\
\multicolumn{4}{c}{c-phase} \\
PBE    & 3.347 & 3.347 & -0.391 \\
r2SCAN & 3.894 & 3.894 & -0.378 \\
HSE06  & 5.042 & 5.042 & -0.379 \\
\multicolumn{4}{c}{f-phase} \\
PBE    & 3.901 & 4.153 & 0.163 \\
r2SCAN & 4.448 & 4.685 & 0.176 \\
HSE06  & 5.626 & 5.910 & 0.204 \\
\multicolumn{4}{c}{ao-phase} \\
PBE    & 3.856 & 4.031 & 0.118 \\
r2SCAN & 4.430 & 4.605 & 0.158 \\
HSE06  & 5.602 & 5.787 & 0.181 \\
\multicolumn{4}{c}{no-phase} \\
PBE    & 3.701 & 3.701 & -0.038 \\
r2SCAN & 4.238 & 4.238 & -0.034 \\
HSE06  & 5.391 & 5.391 & -0.031 \\
\end{tabular}
\end{center}
\end{table}

\begin{table}[!ht]
\begin{center}
\caption{\label{tab:gapsZO}
  Band gaps of ZrO$_2$ phases.
}
\begin{tabular}{lccc}
          & indirect & direct & diff. \\
\hline
\multicolumn{4}{c}{m-phase} \\
PBE    & 3.610 & 3.678 & 0.000 \\
r2SCAN & 4.129 & 4.194 & 0.000 \\
HSE06  & 5.278 & 5.348 & 0.000 \\
PBE+\gowo & 5.222 & 5.296 & 0.000 \\
\multicolumn{4}{c}{t-phase} \\
PBE    & 4.017 & 4.113 & 0.407 \\
r2SCAN & 4.578 & 4.679 & 0.449 \\
HSE06  & 5.737 & 5.830 & 0.460 \\
\multicolumn{4}{c}{c-phase} \\
PBE    & 3.306 & 3.306 & -0.304 \\
r2SCAN & 3.867 & 3.867 & -0.261 \\
HSE06  & 5.024 & 5.024 & -0.254 \\
PBE+\gowo & 4.823 & 4.823 & -0.399 \\
\multicolumn{4}{c}{f-phase} \\
PBE    & 3.841 & 3.973 & 0.231 \\
r2SCAN & 4.390 & 4.511 & 0.262 \\
HSE06  & 5.559 & 5.704 & 0.281 \\
\multicolumn{4}{c}{ao-phase} \\
PBE    & 3.706 & 3.821 & 0.096 \\
r2SCAN & 4.273 & 4.395 & 0.144 \\
HSE06  & 5.434 & 5.576 & 0.156 \\
\multicolumn{4}{c}{no-phase} \\
PBE    & 3.562 & 3.562 & -0.048 \\
r2SCAN & 4.097 & 4.097 & -0.032 \\
HSE06  & 5.249 & 5.249 & -0.029 \\
\end{tabular}
\end{center}
\end{table}

\begin{figure*}[!h]
\begin{center}
  \includegraphics[width=\textwidth]{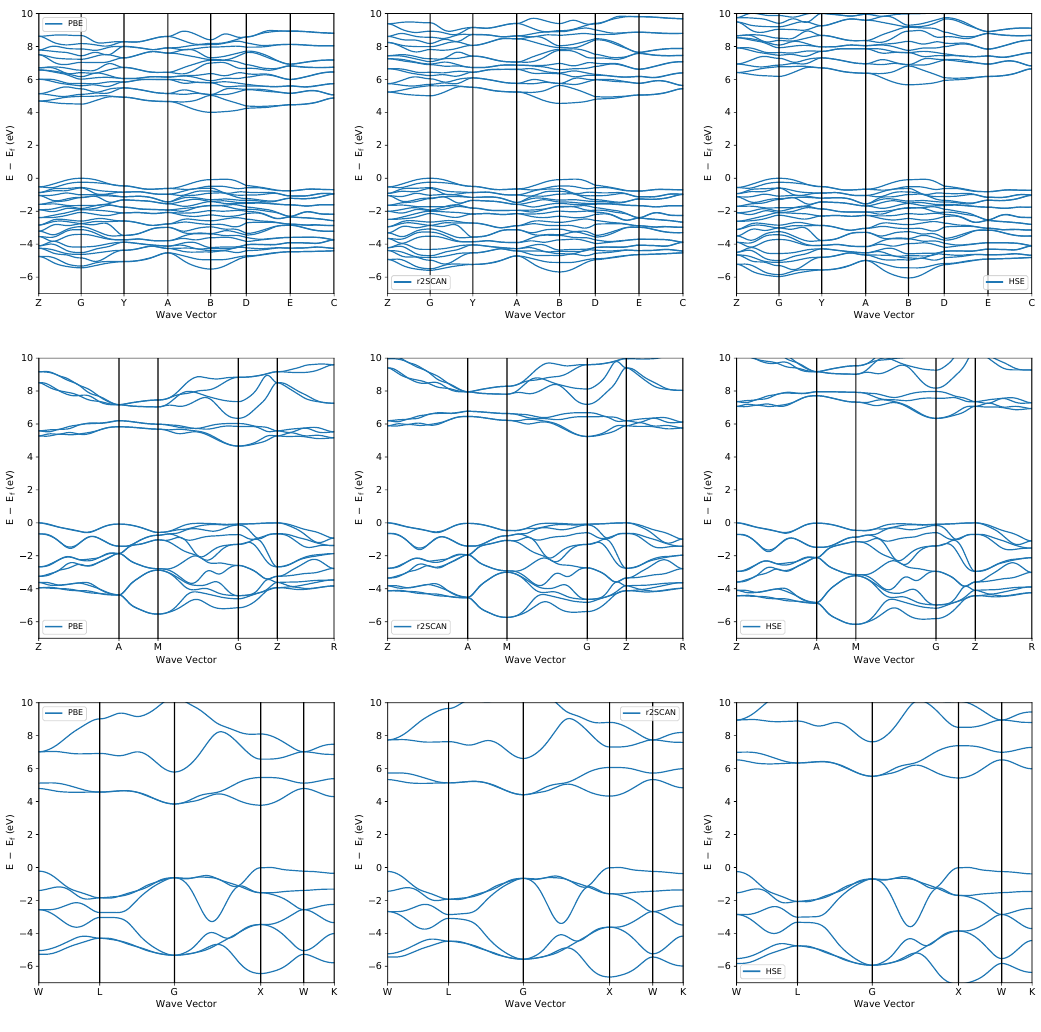}
\end{center}
\caption{\label{fig:bs_ho1}
  Band structures of HfO$_2$ for m-phase (top), t-phase (middle),
  and c-phase (bottom) calculated using PBE, r2SCAN, and HSE06 functionals.
}
\end{figure*}

\begin{figure*}[!h]
  \ContinuedFloat
\begin{center}
  \includegraphics[width=\textwidth]{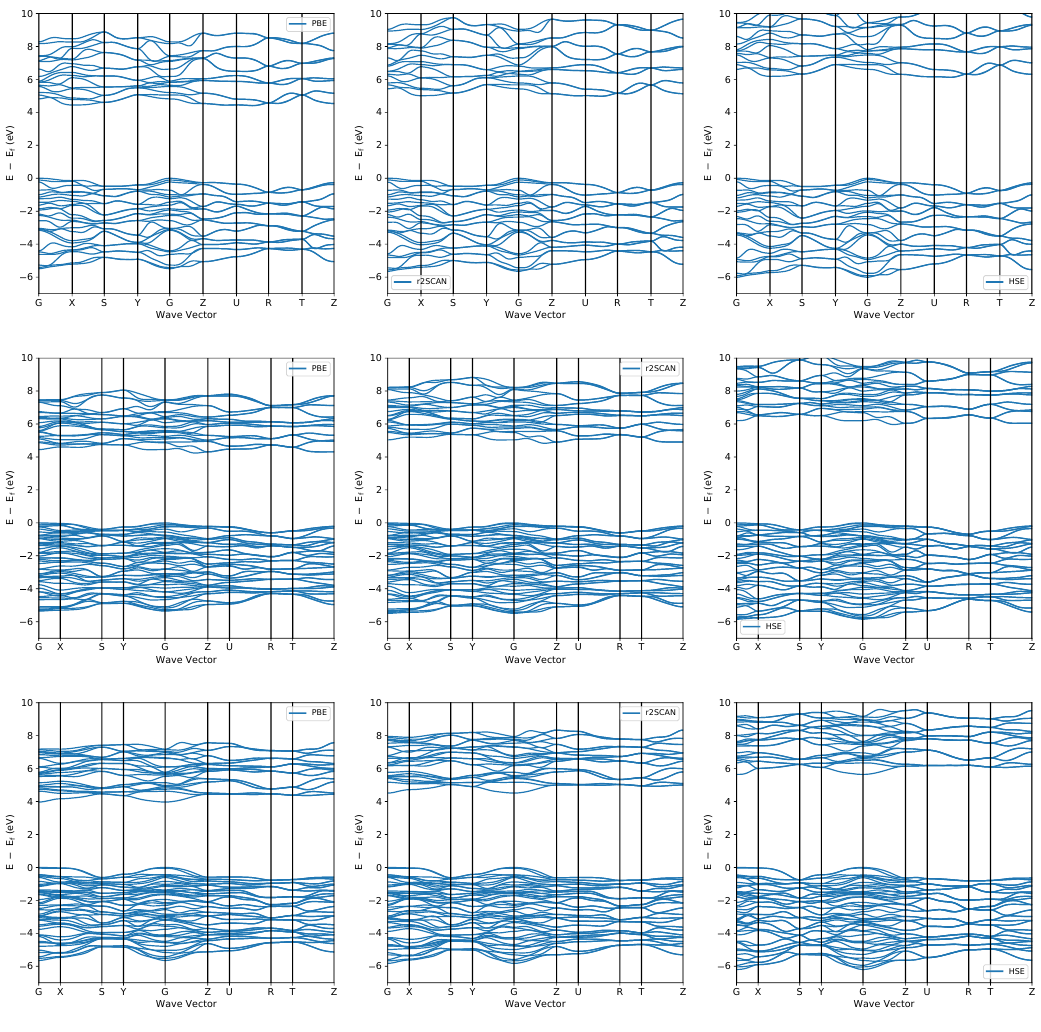}
\end{center}
\caption{\label{fig:bs_ho2} (continued) 
  Band structures of HfO$_2$ for f-phase (top), ao-phase (middle),
  and no-phase (bottom) calculated using PBE, r2SCAN, and HSE06 functionals.
}
\end{figure*}

\begin{figure*}[!h]
\begin{center}
  \includegraphics[width=\textwidth]{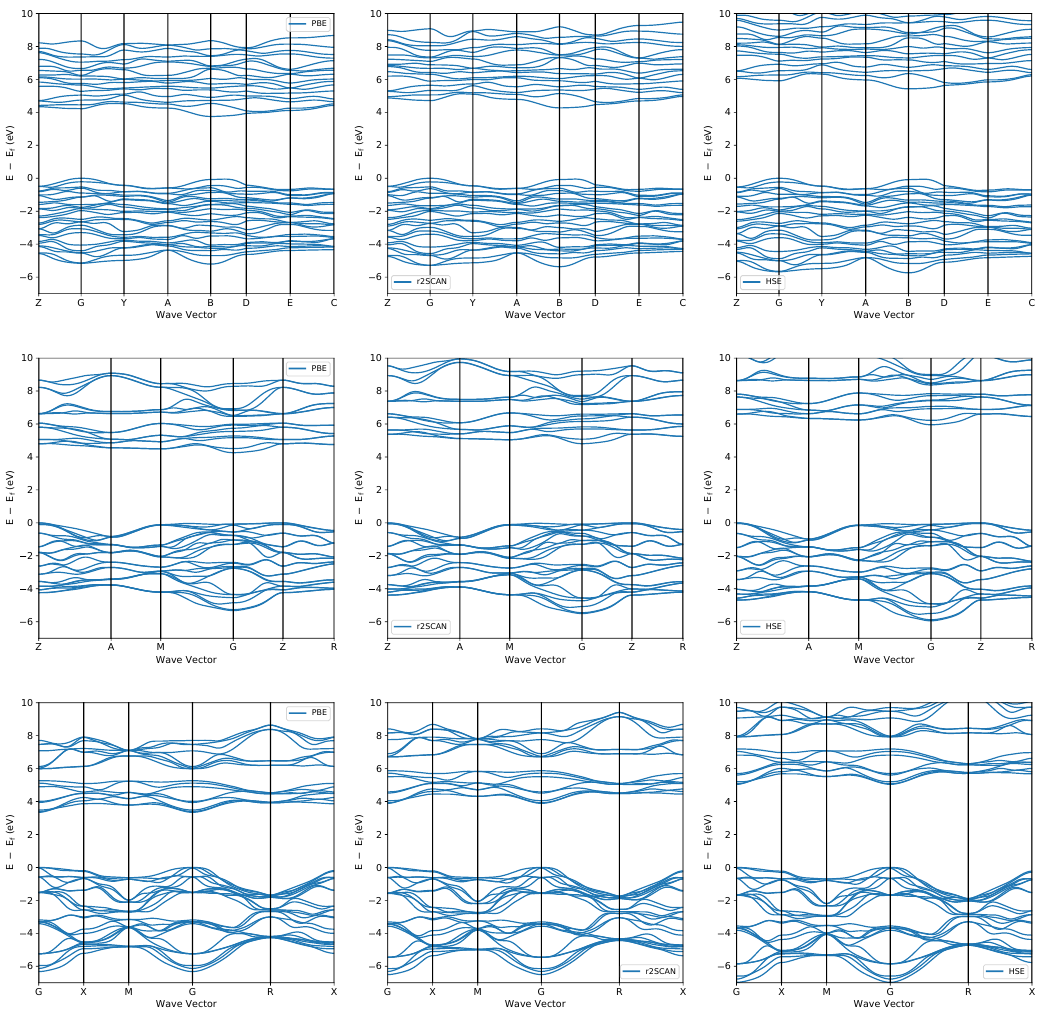}
\end{center}
\caption{\label{fig:bs_hzo1}
  Band structures of {\hzo} for m-phase (top), t-phase (middle),
  and c-phase (bottom) calculated using PBE, r2SCAN, and HSE06 functionals.
}
\end{figure*}

\begin{figure*}[!h]
  \ContinuedFloat
\begin{center}
  \includegraphics[width=\textwidth]{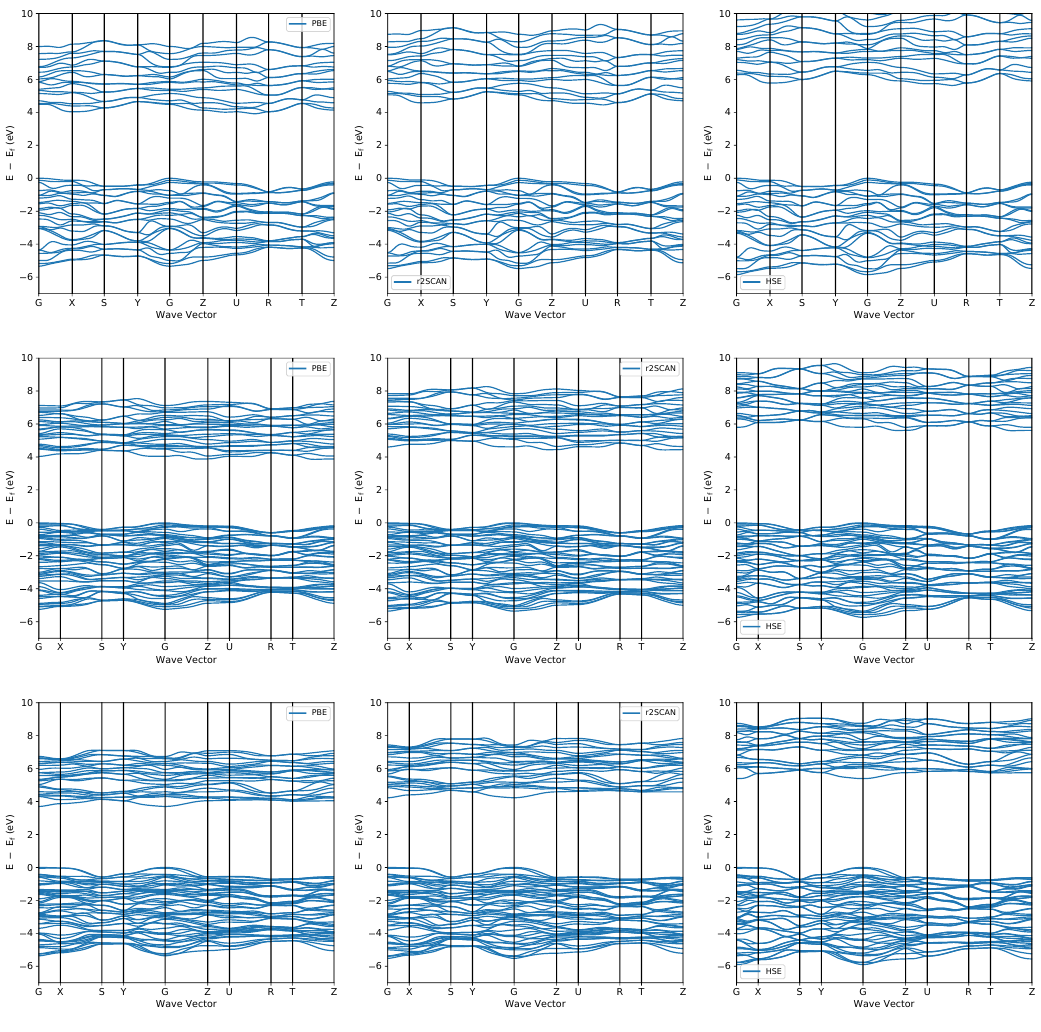}
\end{center}
\caption{\label{fig:bs_hzo2} (continued) 
  Band structures of {\hzo} for f-phase (top), ao-phase (middle),
  and no-phase (bottom) calculated using PBE, r2SCAN, and HSE06 functionals.
}
\end{figure*}

\begin{figure*}[!h]
\begin{center}
  \includegraphics[width=\textwidth]{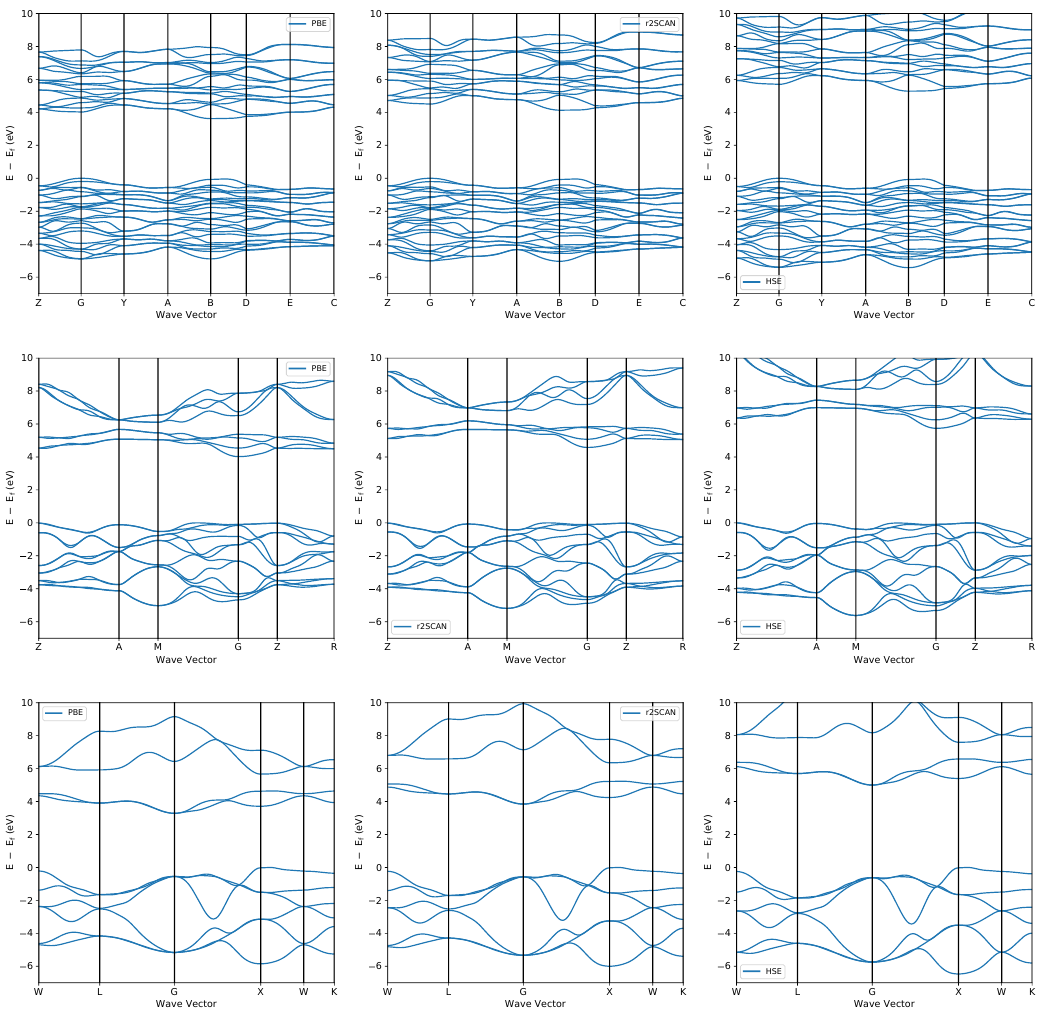}
\end{center}
\caption{\label{fig:bs_zo1}
  Band structures of ZrO$_2$ for m-phase (top), t-phase (middle),
  and c-phase (bottom) calculated using PBE, r2SCAN, and HSE06 functionals.
}
\end{figure*}

\begin{figure*}[!h]
  \ContinuedFloat
\begin{center}
  \includegraphics[width=\textwidth]{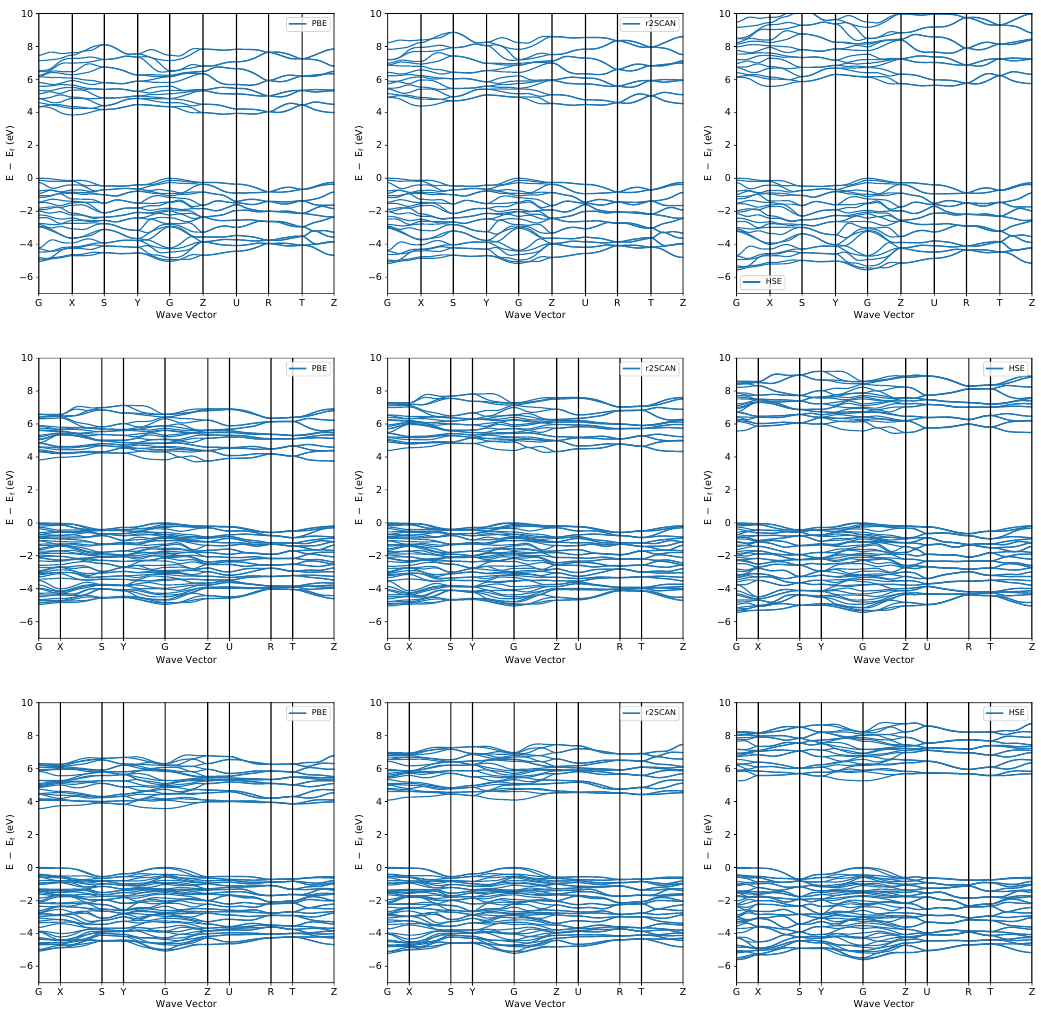}
\end{center}
\caption{\label{fig:bs_zo2} (continued) 
  Band structures of ZrO$_2$ for f-phase (top), ao-phase (middle),
  and no-phase (bottom) calculated using PBE, r2SCAN, and HSE06 functionals.
}
\end{figure*}

\clearpage

\bibliography{refs.bib}